\documentclass[twocolumn]{aastex63}
\usepackage{natbib,enumerate,graphicx,mathtools,epstopdf}
\usepackage[normalem]{ulem}
\usepackage{color,url}
\usepackage{amsthm,amsmath,amsfonts,amssymb}
\newcommand{\powunits}{\mu\textrm{K}^2\ h^{-3}\,\textrm{Mpc}^{3}}
\newcommand{\lineunits}{\mu\textrm{K}^2\ \textrm{Hz}\ \textrm{sr}}

\defcitealias{Li2016}{L16}
\defcitealias{Keating2016}{K16}
\received{July 2, 2020}
\revised{August 7, 2020}
\accepted{August 17, 2020}
\submitjournal{ApJ}

\begin{document}
\title{An Intensity Mapping Detection of Aggregate CO Line Emission at 3 mm}

\correspondingauthor{Garrett K. Keating}
\email{garrett.keating@cfa.harvard.edu}

\author[0000-0002-3490-146X]{Garrett K. Keating}
\affiliation{Center for Astrophysics $|$ Harvard \& Smithsonian, 60 Garden Street, Cambridge, MA 02138, USA}
\affiliation{Academia Sinica Institute of Astronomy and Astrophysics, 645 N. A'ohoku Pl., Hilo, HI 96720, USA}

\author[0000-0002-2367-1080]{Daniel P. Marrone}
\affiliation{Steward Observatory, University of Arizona, 933 North Cherry Avenue, Tucson, AZ 85721, USA}

\author[0000-0003-4056-9982]{Geoffrey C. Bower}
\affiliation{Academia Sinica Institute of Astronomy and Astrophysics, 645 N. A'ohoku Pl., Hilo, HI 96720, USA}

\author[0000-0003-1859-9640]{Ryan P. Keenan}
\affiliation{Steward Observatory, University of Arizona, 933 North Cherry Avenue, Tucson, AZ 85721, USA}

\shorttitle{mmIME I}
\shortauthors{Keating et al.}
\begin{abstract}
We present a detection of molecular gas emission at $z\sim1-5$ using the technique of line intensity mapping. We make use of a pair of 3 mm interferometric data sets, the first from the ALMA Spectroscopic Survey in the Hubble Ultra Deep Field (ASPECS), and the second from a series of Atacama Compact Array (ACA) observations conducted between 2016 and 2018, targeting the COSMOS field. At 100 GHz, we measure non-zero power at 97.8\% and 99.9\% confidence in the ACA and ALMA data sets, respectively. In the joint result, we reject the zero-power hypothesis at 99.99\% confidence, finding $\tilde{I}^{2}_{s}(\nu)=770\pm210\ \lineunits$. After accounting for sample variance effects, the estimated spectral shot power is $\tilde{I}^{2}_{s}(\nu)=1010_{-390}^{+550}\ \lineunits$. We derive a model for the various line species our measurement is expected to be sensitive to, and estimate the shot power to be $120_{-40}^{+80}\ \powunits$, $200^{+120}_{-70}\ \powunits$, and $90^{+70}_{-40}\ \powunits$ for CO(2-1) at $z=1.3$, CO(3-2) at $z=2.5$, and CO(4-3) at $z=3.6$, respectively. Using line ratios appropriate for high-redshift galaxies, we find these results to be in good agreement with those from the CO Power Spectrum Survey (COPSS). Adopting $\alpha_{\rm CO}=3.6\ M_{\sun}\ (\textrm{K}\ \textrm{km}\ \textrm{s}^{-1}\ \textrm{pc}^{2})^{-1}$, we estimate a cosmic molecular gas density of  $\rho_{\textrm{H}_2}(z)\sim 10^{8}\ M_{\sun}\ \textrm{Mpc}^{-3}$ between $z=1-3$.
\end{abstract}

\section{Introduction}\label{sec:intro}
Tracking the evolution of cold molecular gas over cosmic time has become an active research area in recent years, as the amount of observational data on the molecular gas content of high-redshift galaxies has increased dramatically. Facilities like the Submillimeter Array (SMA; \citealt{Ho2004}), the Combined Array for Research in Millimeter-wave Astronomy (CARMA; \citealt{Bock2006}), the Northern Extended Millimeter Array (NOEMA; \citealt{Guilloteau1992}), the Atacama Large Millimeter Array (ALMA; \citealt{Wootten2003}), and the Very Large Array (VLA; \citealt{Heeschen1975}), as well as many others, have increased the number of detections from only a few to thousands over the past two decades \citep{Carilli2013,Tacconi2020}. However, most of these measurements are of galaxies which have been optically selected, or are otherwise significantly more luminous than the majority of the star-forming population. Both of these effects may bias the resulting estimates of the high-redshift molecular gas history.

Blind detections of more common galaxies remains an observationally expensive task, with narrow-field surveys only yielding a dozen or so high-fidelity detections of CO in early galaxies \citep{Walter2014,Pavesi2018,GonzalezLopez2019}. Large sky-area surveys capable of performing volume-limited blind surveys at high redshift will likely require significant advancement in instrumentation, with next-generation facilities that are potentially decades away from operation (e.g., \citealt{Murphy2017}). As a result, there are variations of an order of magnitude in observational constraints and theoretical estimates for the cosmic molecular gas density fueling early star formation at $z\approx3$ \citep{Keating2016,Riechers2019,Decarli2019,Obreschkow2009c,Sargent2014,Popping2014,Lagos2015,Popping2017}.

A growing number of experiments have sought to use power spectrum methods -- like those utilized for measurements of the cosmic microwave background (CMB; e.g. \citealt{Kaiser1982,Meyer1991,Strukov1992,Smoot1992}) -- for measuring aggregate emission coming from a multitude of galaxies over large volumes. This technique, which is colloquially referred to as ``line intensity mapping'' (see \citealt{Kovetz2017} for a review of the subject), can provide an efficient method for probing faint objects. In line intensity mapping, aggregate emission of galaxies is captured as fluctuations in specific intensity over a three-dimensional volume, where the frequency axis is used as a separate spatial dimension. These fluctuations are characterized by a power spectrum, which measures the variance in intensity as a function of spatial scale. The power spectrum is nominally divided into two regimes: the clustering regime, which is sensitive to large-scale structure, and the shot regime, in which the random distribution of galaxies dominate. Unlike direct detection experiments, where the limiting factor is typically point-source sensitivity, the driving figure of merit for intensity mapping is surface brightness sensitivity, which can be achieved with small aperture telescopes. For the cool-gas tracers of CO and [CII], significant theoretical work has been performed in the past decade (e.g., \citealt{Righi2008,Visbal2011,Lidz2011,Breysse2014,Mashian2015,Padmanabhan2018,Yue2019,Sun2019}), and a full swath of pathfinder experiments have begun to move forward in earnest (e.g., \citealt{Pullen2013,Crites2014,Uzgil2014,Bower2015,Li2016,Lagache2018}).

One such experiment was the CO Power Spectrum Survey (COPSS). COPSS utilized an existing, well-characterized instrument (the Sunyaev-Zel'dovich Array -- a subset of CARMA) to provide early constraints on the CO(1-0) power spectrum at at 1 cm (27-35 GHz; $z\approx2.6$). COPSS was divided into two separate stages. The first stage (COPSS I; \citealt{Keating2015}) made use of an older, previously published published data set \citep{Sharp2010}. The second stage (COPSS II; \citealt{Keating2016}, hereafter referred to as \citetalias{Keating2016}) utilized data from a survey optimized for intensity mapping. The final results of \citetalias{Keating2016} reported a tentative detection of the CO(1-0) power spectrum, having measured the aggregate line emission over a volume of $\sim10^{7}\ \textrm{Mpc}^{3}$. Shortly thereafter, \cite{Pullen2018} reported a similar detection of aggregate line emission from [CII] at $z\approx2.6$, having extracted the signal in cross-correlation of the \textit{Planck} 545 GHz intensity map, and \cite{Uzgil2019} placed upper limits on CO line emission utilizing data collected at 3-mm (84-115 GHz) with ALMA. While the results of these three experiments require further observational follow-up, they demonstrate the capabilities of existing instruments and data for placing preliminary intensity mapping constraints.

Following the publication of the COPSS results, we now seek to further explore the potential of intensity mapping experiments to probe high-redshift molecular gas. Within the 3-mm atmospheric window, there are several lines that we expect to be luminous for a Milky Way-like galaxy -- e.g., CO(2-1), CO(3-2), and CO(4-3) -- which are accessible for $z>1$ galaxies \citep{Dannerbauer2009,Daddi2015}. Therefore, intensity mapping analyses targeting this window could offer a relatively low-cost method for detecting molecular gas emission from high-redshift galaxies. The experiment presented here, which we refer to as the Millimeter-wave Intensity Mapping Experiment (mmIME), utilizes a combination of archival and targeted observations. The work presented here is focused on this 3-mm window, with the intention of providing initial constraints on the power spectrum of lines in the millimeter-regime, building upon the results of COPSS, probing the shot regime of the power spectrum.

We have structured this paper in the following way. We describe the data used for our analysis in Section~\ref{sec:data}. We discuss the analysis methods and data verification tests used in processing the data in Section~\ref{sec:analysis}, with the results of this analysis found in Section~\ref{sec:results}. These results are then discussed in further detail in Section~\ref{sec:disc}, with conclusions presented in Section~\ref{sec:conclusion}. Where appropriate in this paper, we assume a $\Lambda$CDM cosmology, with $h=0.7$, $\Omega_{\textrm{m}}=0.27$, $\Omega_{\Lambda}=0.73$. Unless otherwise noted, throughout this paper we report uncertainties in terms of the 68.3\% confidence interval (i.e., $\pm1\sigma$).

\section{Data}\label{sec:data}
The data used for this analysis originate from two separate projects. One of these projects is the ALMA Spectroscopic Survey Hubble Ultra Deep Field Large Program (ASPECS LP), described further in Section~\ref{ssec:data_aspecs}. The other project utilized the Atacama Compact Array (ACA; \citealt{Iguchi2009}), which are described in further detail in Section~\ref{ssec:data_aca}. 

\subsection{ASPECS Data}\label{ssec:data_aspecs}
A more through description of the ASPECS can be found in \cite{GonzalezLopez2019} and \cite{Decarli2019}, though for the sake of completeness, we include a brief description of the project here.

ASPECS is a large-scale program that utilized deep ALMA integrations to measure line and continuum emission from high-redshift galaxies, using data collected at both 3 mm and 1.3 mm. As we have not utilized the 1.3 mm data in our analyses, we will hereafter only discuss the 3 mm observations, and throughout this paper will refer to the ASPECS LP 3 mm data set as the ``ASPECS data''. The 3 mm observations were conducted during ALMA Cycle 4, in December 2016. Five separate tunings were used to cover the entirety of Band 3, which spans from 84-115 GHz, with partially overlapping tunings providing added sensitivity between 96 and 104 GHz. The target of these observations was a 4.6 arcmin$^2$ region within \textit{Hubble} Ultra Deep Field \citep{Beckwith2006}, centered on the position $\alpha=03^{\rm h}32^{\rm m}38.5^{\rm s}$, $\delta=-27^{\circ}47^{\prime}00^{\prime\prime}$ (J2000). The region in question was covered with a 17-point mosaic, with typical dwell times of 0.5 hours per pointing for each tuning, over the course of $\sim50$ tracks total. All data were recorded at 3.906 MHz spectral resolution, with two spectral windows of 1.875 GHz apiece per sideband. Data were collected in the configurations C40-3 and C40-4, with baselines of between 15-460 m and 15-704 m respectively.

\subsection{ACA Data}\label{ssec:data_aca}
\begin{figure*}[t]
    \centering
    \includegraphics[scale=0.37]{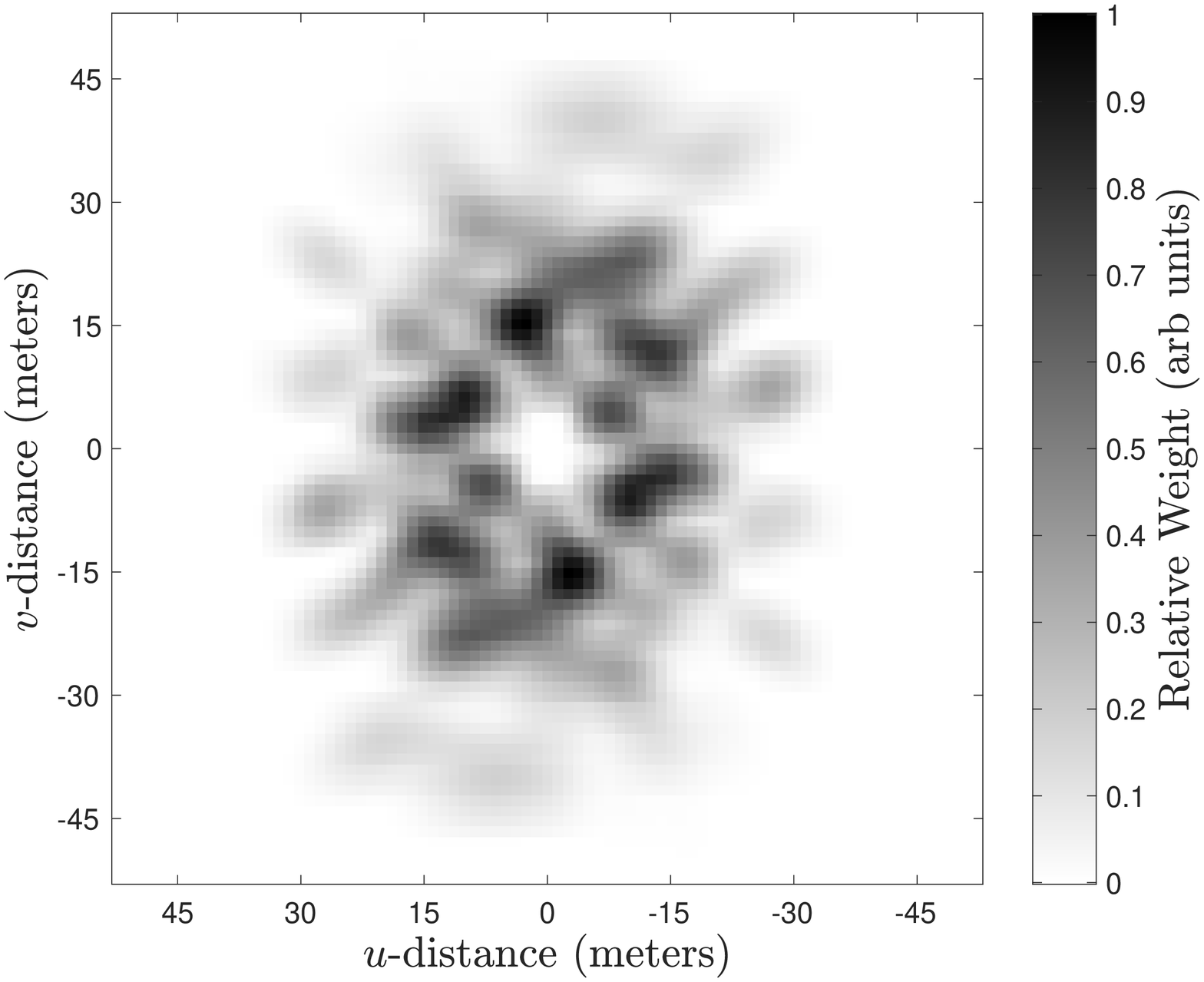}
    $\quad$
    \includegraphics[scale=0.37]{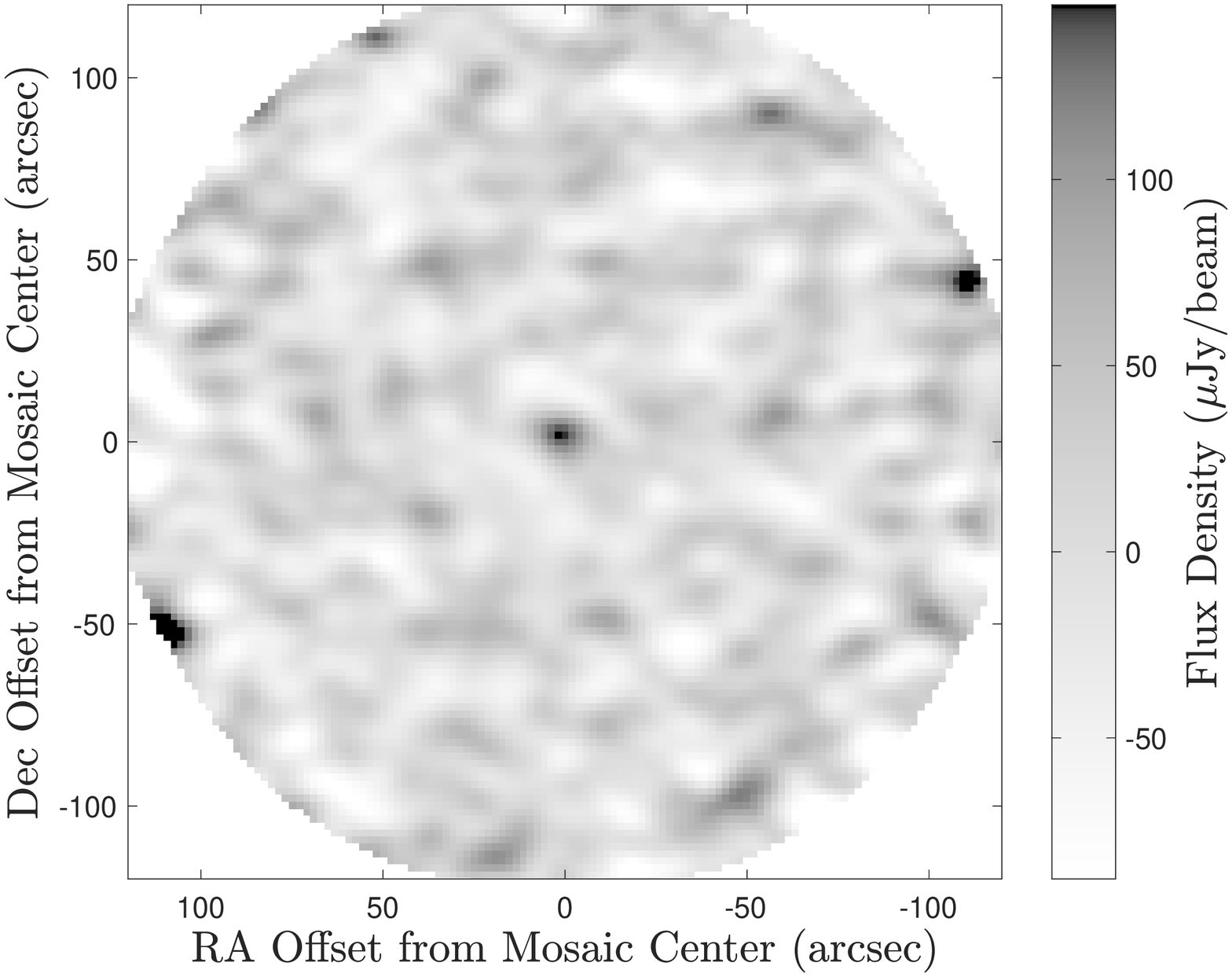}
    \caption{\emph{Left}: The combined $uv$-coverage of the central mosaic pointing of the COSMOS field, from data collected during the 2016.1.01149.S project (see Table~\ref{table:datasets} for more details). The $uv$-plane extent of each visibility weight is determined from the cross-correlation of the illumination patterns of the ACA antennas, which reveals the full range of spatial scales sampled at a $uv$ point. The color scale represents the inverse variance weight of the data at a given position in the $uv$-plane, where darker shading indicates greater weighting. \emph{Right}: The mosaicked continuum image from the entirety of the ACA observations, corrected for primary-beam effects, out to the half-power point of the mosaic pattern. We find a flux density of $150\pm30\ \mu\textrm{Jy}\ \textrm{beam}^{-1}$ at the position of AzTEC-3, in good agreement with previously published results \citep{Riechers2020}.
    \label{fig:aca_img}}
\end{figure*}

The ACA, also known as the Morita Array, is a compact interferometer in northern Chile, consisting of twelve 7-m antennas, adjacent to ALMA. It shares much of the same hardware with the larger ALMA array, with the exception of a dedicated FX-correlator, with 1.75 GHz of usable bandwidth per spectral window, and 2 simultaneous spectral windows per sideband recorded during observations. Unlike the larger 12-m array, the ACA remains in a fixed configuration over time, with baselines between 7 and 45 m in length, as shown in Figure~\ref{fig:aca_img}.

For these observations, we focused on an area of sky within the COSMOS field \citep{Scoville2007}, centered on the position of a source known as AzTEC-3 ($\alpha=10^{\rm h}00^{\rm m}20.7^{\rm s}$, $\delta=+02^{\circ}35'17''$, J2000), an ultra-luminous infrared galaxy (ULIRG) residing at $z=5.3$ \citep{Scott2008,Dwek2011}, which has also been targeted by the VLA and SMA for for blind searches for CO emission at high redshift (\citealt{Pavesi2018}, Keating et al. in prep). A survey area of 15 arcmin$^2$ was covered using a 19-point hexagonally packed mosaic, with pointings spaced such that the field was Nyquist sampled at any given frequency within Band 3 (84-115 GHz). Our goal for these observations was to efficiently probe aggregate emission in the 3 mm atmospheric window, using as few tunings as possible and avoiding frequencies with greater atmospheric contributions to the noise (e.g. close to the 118 GHz O$_2$ line). A total of three tunings were used to cover the spectral window of interest, with data recorded at 7.813 MHz spectral resolution. The first tuning was observed during Cycle 4 (Project ID 2016.1.01149.S) between October 2 and November 10, 2016. The latter two tunings were observed during Cycle 6 (Project ID 2018.1.01594.S), between October 3 and December 11, 2018. Dwell times on each pointing of the mosaic were approximately 2 hours per tuning, observed over the course of $\sim60$ tracks tracks between Cycle 4 and Cycle 6.

\begin{deluxetable*}{c|cccc}
\tablecaption{Summary of data sets used in the analysis presented in this paper
\label{table:datasets}}
\tablehead{
\colhead{Project Code} & \colhead{Array} & \colhead{Freq Coverage} & \colhead{Effective Sky Area$^a$} & \colhead{Integration Time} \\
\colhead{ } & \colhead{ } & \colhead{$[$GHz$]$} & \colhead{$[$arcmin$^2]$} & \colhead{$[$hours$]$}
}
\startdata
2016.1.00324.L & 12-m & $84-115$ & 2.9 & 38.1 \\
2016.1.01149.S & 7-m & $94-98$, $106-110$ & 9.7 & 40.2 \\
2018.1.01594.S & 7-m & $86-94$, $98-106$ & 9.8 & 80.6 \\
\enddata
\tablenotetext{a}{Sky area is calculated as $\Omega_B^{2}=\int{A^{2}\,d\Omega}$, where $A$ is the windowing function over the sky area created by the mosaicking pattern of the relevant observations.}
\end{deluxetable*}

\section{Analysis}\label{sec:analysis}
\begin{figure}
    \centering
    \includegraphics[scale=0.5]{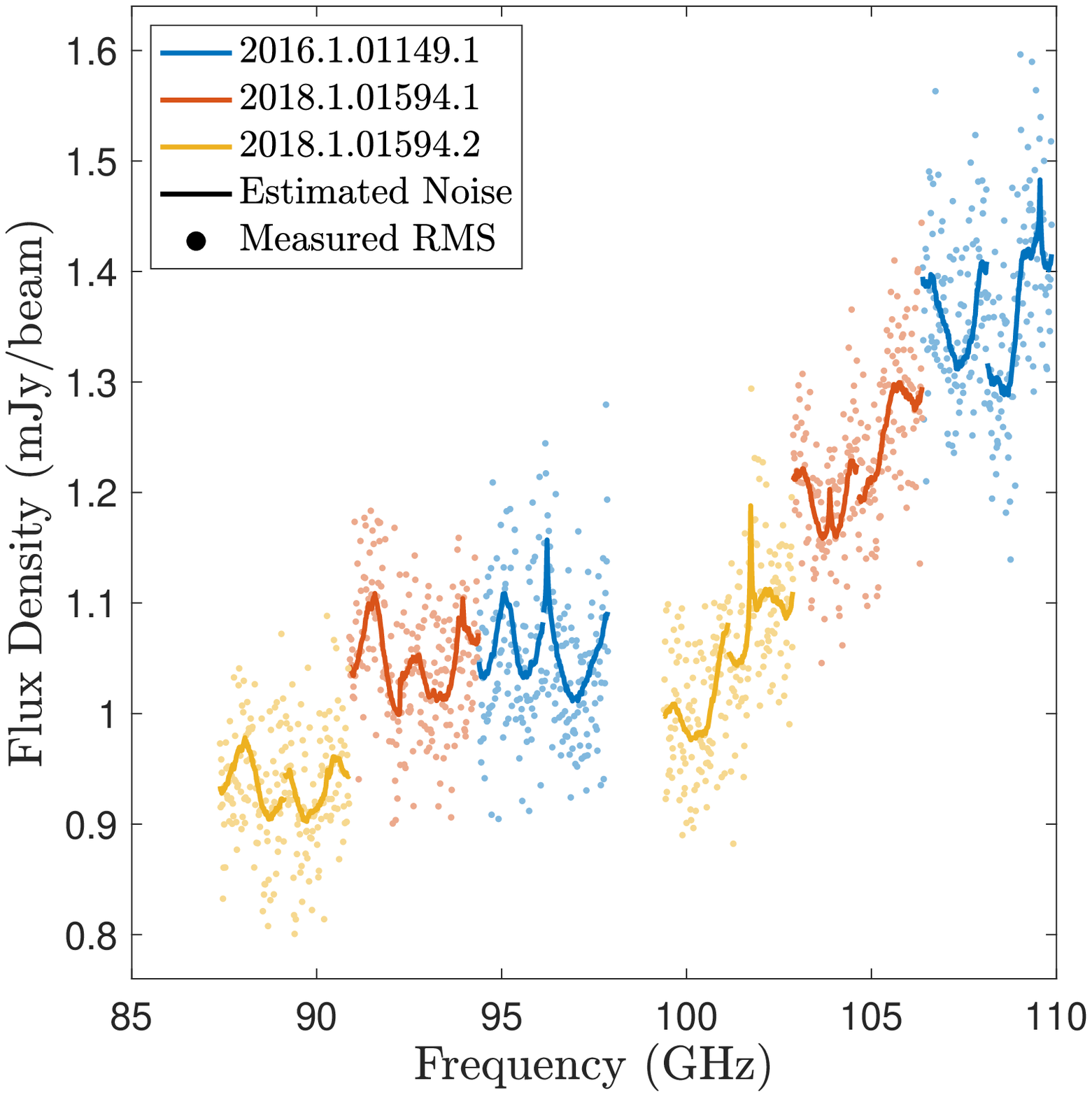}
    \caption{Estimated and measured noise as a function of frequency for the ACA data set. The measured noised is derived from the RMS of the inner quarter of mosaicked image, as calculated per 15.625 MHz channel. We find excellent agreement between the predicted and measured values, indicating that the noise properties of the data are well-characterized following the application of the SEFD corrections derived by our pipeline. The image noise appears to have a smooth, periodic structure as a function of frequency, the presence of which is not natively captured by the \textit{CASA} pipeline. Some small peaks in noise can be seen at 96.2, 101.7, and 109.6 GHz, resulting from absorption by stratospheric ozone. 
    \label{fig:img_noise}}
\end{figure}

Analysis of the data closely follows the prescription used for \citetalias{Keating2016}. We briefly describe our analysis process, noting in detail where the analysis presented here differs from that of previous works, and direct the reader to \citetalias{Keating2016} and \citet{Keating2015} for further details on the analysis pipeline.

\subsection{Data Reduction Pipeline}\label{sec:analysis_pipeline}
For all datasets, we utilize the existing ALMA pipeline software \citep{Shinnaga2015} within the \textit{Common Astronomy Software Applications} package (CASA; \citealt{Jaeger2008}) to provide first-order gains and bandpass corrections, flux scaling, as well as noise estimates. The calibrated visibilities and associated metadata are then exported from CASA into our MATLAB\footnote{Version 2019a, \url{www.mathworks.com}.}-based reduction software package.

After data export, the data from each individual scheduling block are analyzed in their native spectral resolution. An initial round of flagging is performed, which looks for outliers in amplitude when coherently averaging over increasing time intervals for each channel within each baseline. Bandpass solutions are then re-derived and applied to the dataset, with typical corrections of order 1\% in amplitude, and 1 degree in phase.

To better characterize the noise properties of the data, we difference visibilities that are adjacent in time (e.g., for a given baseline and spectral channel, the first integration is differenced with the second, and so on), such that the resultant dataset is expected to contain only noise. With this noise-only dataset, we perform two tasks. First, we fit for antenna-based channel-by-channel system equivalent flux density (SEFD) corrections, which help account for IF or quantization losses that may not be captured by system temperature or efficiency measurements \citep{Keating2015}. Second, we measure the covariance of the noise for all channels within a given spectral window, averaged across all baselines. With the noise properties measured, we average down the data to 15.625 MHz resolution for both ACA and ASPECS data. For both data sets, we find that there are moderate frequency-dependent SEFD corrections required. These corrections are largest at the edges of spectral windows, which may be result of increased quantization noise in low-amplitude channels, requiring SEFD corrections of order $10{-}20\%$ above that estimated by CASA. These corrections appear relatively consistent in data collected within a single ALMA Cycle. The net impact of these corrections can be seen in the agreement between the estimated and measured noise in the image domain, as shown for the ACA data set in Figure~\ref{fig:img_noise}.

Additional flagging is performed in the spectrally averaged data. The flags are determined from coherent and incoherent sums of data across time intervals from 6 seconds to $\sim$2 hours, evaluated on a per-channel, per-window, and per-baseline basis. This typically discards $3{-}5\%$ of data on top of what the CASA pipeline already flags, with the exception of a single track each from the ACA and ASPECS data, where our analysis flagged closer to $20{-}30\%$ of data. We discard all data from these two tracks, amounting to a loss of $2\%$ of the total data volume. Bandpass calibration, antenna-based SEFD correction, and noise covariance are re-derived and checked for consistency against that derived for the full resolution data. Antenna gain corrections are then derived and applied to the data, with typical corrections of order 2-3\% in amplitude, and 1 degree in phase. Where gains are observed to vary by more than 15\% between calibration cycles, the offending antenna is flagged. For both ACA and ASPECS data sets, this is a relatively rare occurrence, affecting less than 0.1\% of all data. Generally speaking, we find that both the bandpass and antenna gains shown require only modest corrections after processing through the ALMA pipeline, such that the relative errors on these quantities are expected to of order less than 1\%, and are not expected to be a limiting factor in our analysis. The absolute flux uncertainty is estimated to be of order $\sim$5\%, although we note that this number has been subject to debate within the ALMA community \citep{Bonato2018}. Nevertheless, we expect this uncertainty to be small relative to the overall uncertainty our measurement, barring a high-significance detection ($\gtrsim10\sigma$).

The data from each scheduling block are then imaged, solely for diagnostic purposes. We show the resulting continuum image from the full set of ACA observations in Figure~\ref{fig:aca_img}. When imaging the ASPECS data (not shown), we find good agreement with the imaging results published in \citet{GonzalezLopez2019} for the ASPECS data.

After imaging and inspection, the data are then gridded for power spectrum analysis. Each spectral window in each baseline is separately Fourier transformed along the frequency axis to return to the lag domain and produce what we refer to as ``delay-visibilities''. The delay-visibilities are then gridded into a 4-dimensional grid, based on coordinates in $(u,v,\eta,\nu_{\rm c})$, where $u$ and $v$ are the standard interferometer spatial frequency coordinates, measured in units of physical distance (i.e., meters), $\eta$ is the Fourier dual to the frequency axis (i.e., the lag coordinate), measured in inverse frequency units of $\nu_{\rm c}^{-1}$, and $\nu_{\rm c}$ is the center frequency of the spectral window that is being gridded. In order to maintain maximum sensitivity, the data are naturally weighted and gridded using a simple, rectangular (i.e., Shah) gridding function, into cells of size $10\%$ of the antenna diameter. This choice of gridding function helps to maximize processing speed, and ensures that the noise within adjacent cells is uncorrelated, at the cost of $\approx1{-}2\%$ of our final sensitivity. The sub-pixel mean $uv$-position of the data are also recorded, as well as the estimated noise and windowing functions, the latter of which are determined by the measured channel-to-channel noise covariance.

At the completion of processing for an individual observing block, the gridded data are accumulated across all tracks, and power spectrum analysis is conducted on the aggregate data.

\subsection{Power Spectrum Analysis}\label{ssec:powspec_methods}
Throughout this section, we discuss how measurements of the intensity, $I$, and its Fourier dual, $\tilde{I}$, are related to the power spectrum measurements that we seek to perform. While previously derived in \cite{Keating2015}, our methodology has changed to accommodate the mosaicked data sets used here, versus the single pointings used in COPSS, as has the manner in which we have reported the results (for reasons discussed below). Because of this, we present here our power spectrum methodology in full, noting that both $I$ and $\tilde{I}$ are expressed in brightness temperature units. 

The native output of the interferometer is the visibility, $\mathcal{V}$, representing the Fourier transform of the sky intensity at a given frequency, which can be Fourier-transformed across the frequency axis to produce delay-visibilities, $\tilde{\mathcal{V}}(u,v,\eta,\nu_{c})$, that are functions of angular and frequency wavenumbers (as discussed in Section \ref{sec:analysis_pipeline}), and are directly proportional to $\tilde{I}$ \citep{Bond1998,White1999,Morales2004,Parsons2012}. The immediate goal of our analysis is to measure the variance in intensity over different spatial scales, which by virtue of Parseval's theorem, is equivalent to measuring the power, $\tilde{I}^{2}$, over a set of Fourier modes, where
\begin{multline}\label{eqn:intsq_formal}
 \tilde{I}^{2}(u,v,\eta,\nu_{\rm c}) = \\ \frac{\tilde{\mathcal{V}}^{*}(u,v,\eta,\nu_{\rm c})\cdot\tilde{\mathcal{V}}(u,v,\eta,\nu_{\rm c})}
 {\iiint W(l,m,\nu) \cdot W(l,m,\nu)\ dl\,dm\,d\nu}.
\end{multline}
In Equation~\ref{eqn:intsq_formal}, the term in the denominator serves as a normalization factor, where $W(l,m,\nu)$ is the windowing function of the measurement in question. For a measurement which is uniformly weighted across both the plane of the sky and the spectral axis, the denominator simplifies to $B_{\rm surv}\Omega_{\rm surv}$, where $B_{\rm surv}$ is the bandwidth of the observation and $\Omega_{\rm surv}$ is the sky-area of the survey. 

The typical data product of intensity mapping analysis is the power spectrum, $P(k,z)$, or its dimensionless counterpart, $\Delta^{2}(k,z)$. Both are expressed as functions of wavenumber, $k$, and redshift, $z$, and can be further defined as
\begin{eqnarray}\label{eqn:powspec_formal}
\Delta^{2}(k,z) &\equiv& \frac{k^{3}}{2\pi^{2}} P(k,z), \nonumber \\
&=& \frac{k^{3}}{2\pi^{2}} X^{2}Y \left \langle \tilde{I}^{2}(u,v,\eta,\nu_{\rm c}) \right \rangle_{T(u,v,\eta,z,\nu_{\rm r})=k}.
\end{eqnarray}
In Equation~\ref{eqn:powspec_formal}, $T$ a coordinate mapping function between wavenumber units, from the observed $(u,v,\eta,\nu_{\rm c})$ to the co-moving spatial wavenumber, $k$, at a given redshift, $z$, where 
\begin{equation}\label{eqn:trans_obs_to_comoving}
T(u,v,\eta,z,\nu_{\rm r}) = 2\pi \left (\frac{u^{2}+v^{2}}{X^{2}(z)} + \frac{\eta^{2}}{Y^{2}(z,\nu_{\rm r})} \right )^{1/2}.
\end{equation}
In Equations~\ref{eqn:powspec_formal} and \ref{eqn:trans_obs_to_comoving}, the values of $X$ and $Y$ are used for unit conversion between observed and physical units \citep{Parsons2012}, which can be expressed as
\begin{eqnarray}
X(z) &=& D_{M}(z),\label{eqn:xconv}\\
Y(z,\nu_{\rm r}) &=& \frac{c(1+z)^{2}} {H_{0}E(z)\nu_{\rm r}}\label{eqn:yconv}.
\end{eqnarray}
In Equation~\ref{eqn:xconv}, $D_{M}$ refers to the comoving radial distance. In Equation~\ref{eqn:yconv}, $c$ is the speed of light, $H_{0}$ is the Hubble constant at $z=0$, $E(z)$ is the dimensionless Hubble parameter, and $\nu_{\rm r}$ is the rest frequency of the spectral line under consideration. 

There is an intrinsic assumption in Equation~\ref{eqn:powspec_formal}, in that it effectively requires one to choose a singular redshift (or spectral line) to be the source of the line emission being measured. At 3 mm, we expect multiple lines to have contributions to $\tilde{I}^{2}$, without a simple way to separate individual power spectrum components of each line. In the presence of large-scale structure, one can use the relative anisotropies in the 3D power spectrum to separate out their contributions \citep{Cheng2016}. However, over the wavenumbers considered here, we expect the measurement to be solely sensitive to the small-scale fluctuations driven by individual galaxies, typically referred to as the shot power \citep{Lidz2011,Gong2011}, $P_{\rm shot}(z)$. Barring cross-correlation or exhaustive imaging efforts, we risk making an arbitrary choice.

To side-step this problem, we consider what we refer to as the ``spectral shot power'', $\tilde{I}^{2}_{s}(\nu)$, which can be calculated by summing $\tilde{I}^{2}$ over all wavenumbers with $u$, $v$, and $\eta$ greater than some minimum value, set by the scales as which we expect the contributions of large-scale structure to be minimal ($k\gtrsim1\,\textrm{Mpc}^{-1}$). It describes the minimum variance in intensity (absent resolving out individual galaxies) one expects in a power spectrum measurement, varying only as a function of frequency. 

To measure $\tilde{I}^{2}_{s}(\nu)$, we must first calculate $\tilde{I}^{2}$, which we do with the the following estimator:
\begin{equation}
\tilde{I}^{2}(u,v,\eta,\nu_{\rm c}) = \frac{\sum\limits_{i} \sum\limits_{j}  \left [ \psi_{i,j}\tilde{\mathcal{V}}_{i}^{*} \phi_{i,j} \tilde{\mathcal{V}}_{j} w_{i,j} \right ]  {-} \mathcal{A}_{i}}{\sum\limits_{i} \sum\limits_{j} \left [ w_{i,j}  \right ] - w_{\mathcal{A}_{i}}}. \label{eqn:intsq_est}
\end{equation}
In Equation~\ref{eqn:intsq_est}, we are effectively taking a weighted sum of the set of all delay-visibilities cross-multiplied against one another, to produce an estimate of $\tilde{I}^{2}$ as a function of $u$, $v$, $\eta$, and $\nu_{\rm c}$. $\phi$ is a phase rotation term required when cross-multiplying delay-visibilities with different phase centers -- i.e., data from different mosaic pointings. $\psi$ is a normalization term which accounts for the combined the windowing functions of the $i$-th and $j$-th delay-visibilities, which can be further defined as
\begin{equation}\label{eqn:intsq_norm_factor}
\psi_{i,j} = \left [\iiint (A_{i} *  B_{i}) \cdot (A_{j} * B_{j}) \ dl\,dm\,d\nu \right ]^{-1}{,}
\end{equation}
where $A$ and $B$ are the windowing function in the plane of the sky and across the frequency axis, respectively. For our analysis, we take $B(\nu)=1$ over the frequency window of the measurement (and zero outside of this window), and $A$ is set by the primary beam pattern of the antenna, centered on the mosaic pointing for the given delay-visibility. The correlated product of a given pair of modes is weighted by the function $w_{i,j}$, which can be further defined as
\begin{equation}\label{eqn:powspec_weights}
w_{i,j} =  \frac{C_{i,j}}{\sigma^{2}_{i}\sigma^{2}_{j}},
\end{equation}
where $C_{i,j}$ is the expected signal covariance between modes, and $\sigma^{2}_{i}$ is the estimated noise variance. 

After normalization and weighting, the $\mathcal{A}_{i}$ term in Equation~\ref{eqn:intsq_est} is needed to debias the estimator. This factor is the weighted auto-correlation of all delay-visibilities within a single gridded cell, which removes this bias assuming that the noise in adjacent cells is uncorrelated. We similarly subtract off the sum of all weights for the auto-correlations, $w_{\mathcal{A}_{i}}$, in order to properly normalize $\tilde{I}^2$. 

With our measurements of $\tilde{I}^{2}$ in hand, we are now able to calculate quantity $\tilde{I}^{2}_{s}$ by averaging over the available ensemble of modes, 
\begin{equation}\label{eqn:intsq_est_total}
\tilde{I}^{2}_{s}(\nu) = \frac{\sum\limits_{u,v,\eta} \tilde{I}^{2}(u,v,\eta,\nu_{\rm c})\sigma_{\tilde{I}^{2}}^{-2}(u,v,\eta,\nu_{\rm c})}{\sum\limits_{u,v,\eta} \sigma_{\tilde{I}^{2}}^{-2}(u,v,\eta,\nu_{\rm c})} 
\end{equation}
where $\sigma_{\tilde{I}^{2}}$ is the estimated noise variance of $\tilde{I}^{2}$.

Though division of its power into specific lines requires a model or some other constraint, the spectral shot power represents the following summation:
\begin{equation}\label{eqn:linepow_def}
\tilde{I}^{2}_{s}(\nu) = \sum\limits_{\textrm{all lines}}{\frac{P_{\rm shot,line}(z)}{X(z)^{2}Y(z,\nu_{\rm r,line})}},
\end{equation}
where $P_{\rm shot,line}(z_{\rm})$ is the shot power for a given spectral line at a given redshift, $z$, which is set by the rest frequency of the line in question, $\nu_{\rm r,line}$, and the frequency at which the spectral shot power is being evaluated. The primary advantage of using $\tilde{I}^{2}_{s}(\nu)$ is that it requires no assumptions about the underlying cosmology, nor division of the signal into separate redshift bins -- it is solely based in observable units, which we typically express as $\lineunits$. We therefore find it appropriate for the measurements presented here, and will express our results in units of $\tilde{I}^{2}_{s}(\nu)$, unless otherwise noted.

To verify that our analysis methodology is correct, we run a series of validation tests that simulate a mosaicked observation with both the ACA and ALMA 12-m arrays at 100 GHz in the presence of blank sky, and then again with in the presence of a background population of line emitters with known number density. For both, we run a series of $10^4$ trials separately for simulated versions of the 7-m and 12-m arrays to verify that our noise estimates and power spectrum estimator are correct. In both cases, we find good agreement for both to within $\sim1\%$ percent of expectations, in line with what one would expect given the number of trials run.

In our analysis, we exclude a subset of modes from our measurement that have the potential to resolve out the emission from individual galaxies. We exclude baselines of length $\gtrsim {\rm 100\ k\lambda}$, to avoid resolving sources that are $\lesssim2''$ in diameter, which is expected to cover the population of interest \citep{Tacconi2013}. We similarly exclude modes where $\eta\geq500$ to avoid resolving sources a FWHM linewidth of $\lesssim300\ \textrm{km}/\textrm{sec}$. We discuss in further detail the impact this choice of threshold for $\eta$ in Appendix \ref{app:linewidths}. 

\subsubsection{The Impact of Continuum Foregrounds}\label{ssec:foregrounds}
Compared to the sky at 30 GHz (i.e., the frequency at which COPSS was observed), the continuum foregrounds -- particularly those arising from individual point sources -- are relatively weak. At the angular scales considered in this measurement ($\theta<1'$), for targets well outside of the plane of the Galaxy (such as those studied here), the primary continuum contaminant at 3 mm is expected to arise from the contributions of individual point sources \citep{Reichardt2012,PlanckXI2015}. While these contributions are generally weak, in aggregate they can contribute more power than the line emission of interest.

Rather than attempting to subtract out individual sources, which may introduce its own set of artifacts and systematics, we seek a different method of eliminate the power from these potential contaminants. As these continuum sources are expected to be spectrally smooth, their contributions will primarily contaminate modes where $\eta\sim0$. As \citet{Keating2015} demonstrated, discarding the plane of modes at $\eta=0$ can dramatically reduce the power from continuum sources, even in the presence of moderate gain errors, at the cost of some sensitivity in our power spectrum analysis.

Adopting the 3-mm source count model of \cite{Zavala2018}, we find that by removing the $\eta=0$ modes reduces the contaminant power to $\lesssim0.1\ \lineunits$. We note that without excluding these modes, the mean contaminant power rises to $\sim 200\ \lineunits$ for both the ASPECS and ACA datasets, which is potentially comparable to the power arising from line emission. Even screening out point sources above the imaging threshold of ASPECS ($\sim30\ {\rm \mu Jy}$), there still exists power of order $\sim 50\ \lineunits$. As this is comparable to our estimated error from thermal noise alone, we choose to exclude the $\eta=0$ modes, at the cost of $3\%$ of the final sensitivity of our power spectrum measurement.

\subsubsection{Jackknife Analysis}\label{ssec:jackknife}
\begin{figure*}[t]
    \centering
    \includegraphics[width=0.970\textwidth]{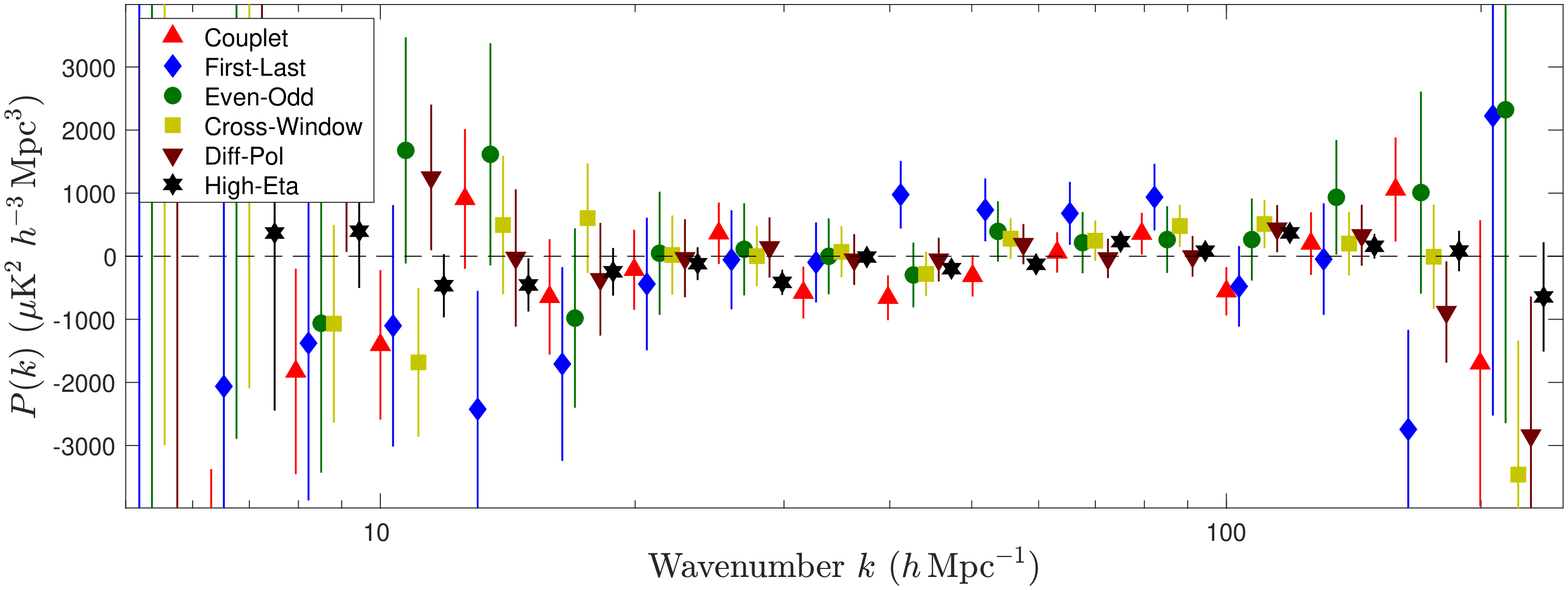}
    \\
    \
    \\
    \includegraphics[width=0.970\textwidth]{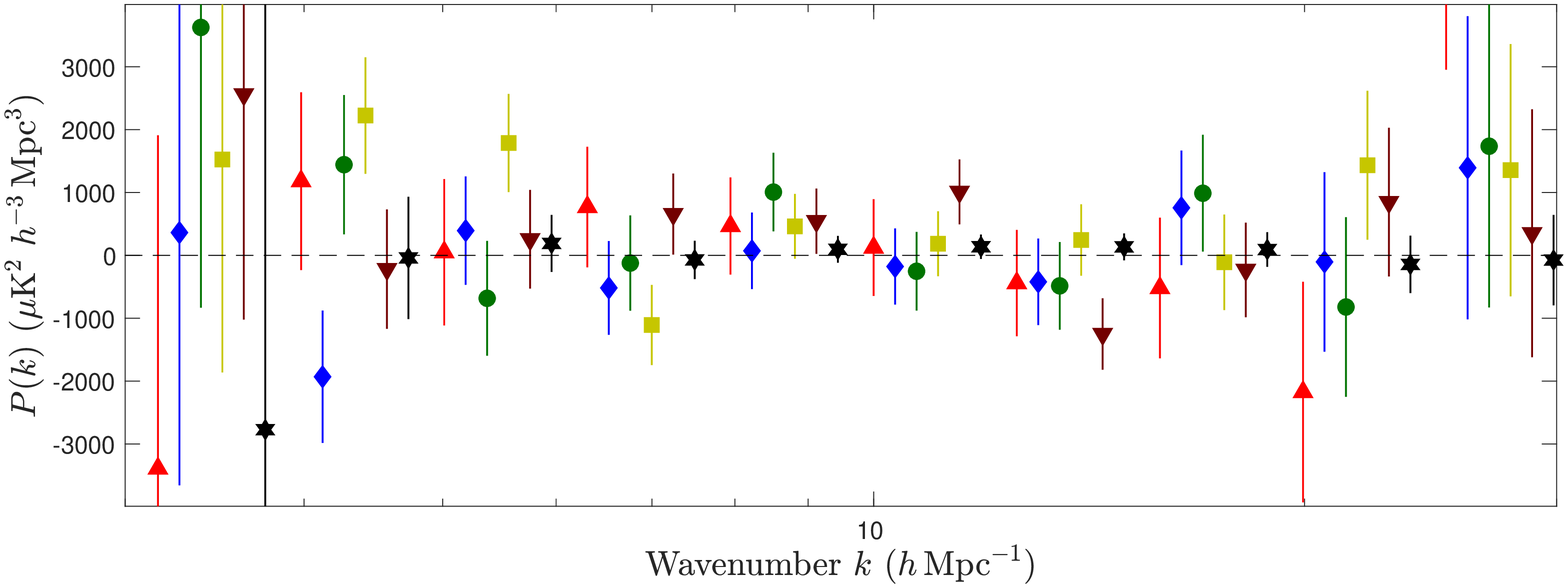}
    \caption{Power measured in our jackknife analyses as a function of wavenumber $k$, projected to the CO(3-2) frame, in units of $\powunits$, for the ASPECS (top panel) and ACA (bottom panel) data sets. The results of the analyses of both data sets are consistent with noise, with the largest outliers observed at $2.4\sigma$ for both the ASPECS and ACA results. Both are consistent with what one would expect for a noise-like distribution, given the number of data points produced in our analysis.
    \label{fig:jackknife}}
\end{figure*}

To verify that our power spectrum measurement is free of significant systematics that may otherwise corrupt our result, we perform a series of jackknife analyses, where the data are combined in such a way that the astrophysical signal in question is subtracted or otherwise made incoherent. We expect the resulting power spectra to be consistent with a null result, so detections of non-zero power are indicative of contaminating signals that are unrelated to the line emission we seek to measure.

We perform six jackknife tests with our data. The following four were previously used in the COPSS survey. The ``couplet'' test differences visibilities that are adjacent in time, subtracting data from one integration from the integration immediately following it. The ``first-last'' test differences the sum of all data taken during the first half of all scheduling blocks from the sum of all data taken during the second half. Similarly, the ``even-odd'' test differences the the sum of all data taken during even-numbered observing blocks and the sum of all odd-numbered observing blocks.  Finally, the ``cross-win'' test correlates modes in different redshift windows, but the same position in $(u,v,\eta)$,  where the spectral lines of interest are not expected to be correlated between different volumes for the modes that we measure (though the same may not be true of continuum contaminants or systematics).

In addition to the previously used jackknife tests, we include two new tests in the analysis presented here. The first test is one we refer to as ``diff-pol'', which leverages the dual-polarization capabilities of ALMA. As the CO signal itself is expected to be unpolarized, we expect that differencing the vertical and horizontal polarization visibilities from one another will remove the signal of interest, but potentially leave behind systematics that are seen preferentially in one polarization over another. The second test is referred to as ``high-$\eta$'', where only modes with $\eta>500$ are included, under the expectation that they resolve out the bulk of the individual sources in frequency-space, leaving little to no power from astronomical sources (discussed in Appendix \ref{sapp:line_profiles}).

\begin{deluxetable*}{c|cc|cc|cc}[t]
\tablecaption{Spectral Shot Power and Jackknife Measurements \label{table:powspec_results}}
\tablehead{
\colhead{ } & \multicolumn{2}{c}{ASPECS Data} & \multicolumn{2}{c}{ACA Data} & \multicolumn{2}{c}{Joint Result$^{\rm a}$} \\
 \colhead{Jackknife Test} & \colhead{Result} & \colhead{\phs PTE} & \colhead{Result} & \colhead{\phs PTE} & \colhead{Result} & \colhead{\phs PTE}}
\startdata
Couplet        & ${-}330\pm250$ & \phs$0.18$        & \phs$310\pm660$ & \phs$0.64$  & ${-}250\pm230$  & \phs$0.28$ \\
First-Last     & \phs$610\pm390$ & \phs$0.12$       & ${-}310\pm520$ & \phs$0.55$   & \phs$280\pm310$ & \phs$0.38$ \\
Even-Odd       & \phs$400\pm380$ & \phs$0.29$       & \phs$330\pm540$ & \phs$0.54$  & \phs$380\pm310$ & \phs$0.22$ \\
Cross-Win      & \phs$300\pm240$ & \phs$0.21$       & \phs$760\pm440$ & \phs$0.09$  & \phs$410\pm210$ & \phs$0.05$ \\ 
Diff-Pol       & \phs$190\pm240$ & \phs$0.44$       & \phs$440\pm440$ & \phs$0.32$  & \phs$250\pm210$ & \phs$0.25$\\
High-$\eta$    & ${-}20\pm\phn100$ & \phs$0.86$      & \phs$170\pm190$ & \phs$0.37$      & $\phn20\pm\phn80$ & \phs$0.78$\\
\hline
Science Result & \phs$730\pm240$ & $<0.01$          & \phs$890\pm440$ & \phs$0.02$  & \phs$770\pm210$ & $<0.01$ \\
\enddata
\tablenotetext{a}{Results calculated using an inverse variance-weighted sum of both data sets.}
\tablecomments{All values are in units of $\lineunits$.}
\end{deluxetable*}

The results from our jackknife analysis are shown in Table~\ref{table:powspec_results} and in Figure~\ref{fig:jackknife}. For both the ASPECS LP and ACA datasets, we find that the results are generally consistent with that which we would expect from noise. Within the results of the ASPECS data, we do find a small excess of power for the first-last jackknife test at $k_{\textrm{CO(3-2)}}\approx50\ h\,\textrm{Mpc}^{-1}$, spread over several separate bins, none of which exceed $2\sigma$ in significance. Optimally weighting over the range between $k_{\textrm{CO(3-2)}}=40{-}90\ h\,\textrm{Mpc}^{-1}$, we measure power at $2.9\sigma$, with the likelihood of a noise-like event to produce equal or greater power (probability to exceed; PTE) equal to approximately 1 in 300. Although not inconsistent with a noise-like event, given the $\sim200$ individual measurements in Figure~\ref{fig:jackknife}, we have explored this result in greater detail. We find no particular redshift or track that dominates the signal. This jackknife is the least sensitive of the six, in part because the slow changes in the configuration of ALMA 12-m array over the course of the observations cause the $uv$-coverage of the first and last halves of the data to be somewhat disjoint, which leaves few measurements per $uv$ cell for differencing. In such cases, the noise distribution becomes skewed \citepalias{Keating2016}, making the likelihood of positive outlier events higher than expected relative to a normal distribution. Evaluating across all wavenumber bins, we find a reduced chi-squared value of $\chi^{2}_{\rm red}=1.04$ (with 18 degrees of freedom), consistent with noise. As our final results are reported as band-averaged values, and as the band-averaged jackknife results for both datasets show less than $2\sigma$ significance, we conclude that the data are sufficiently free of systematics for power spectrum analysis.

As in \citetalias{Keating2016}, we perform one additional jackknife-like test, where the phases of the individual cells in each gridded data set are randomized before accumulation and correlation, in order to further validate our estimated uncertainties. We find that the noise estimates from this test agree with those derived from thermal estimates to within the nominal sample variance once expects from 100 trials (i.e., 10\% uncertainty).

\section{Results}\label{sec:results}
\begin{figure*}
    \centering
    \includegraphics[width=0.975\textwidth]{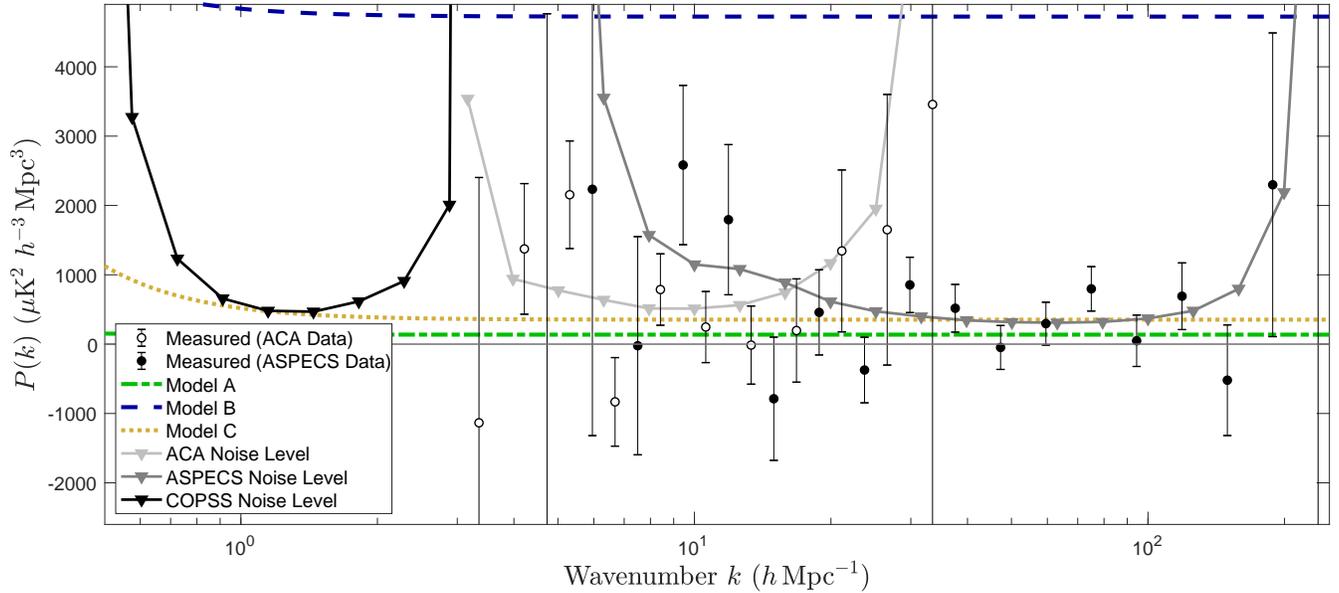}
    \caption{Power measured in our primary analysis of the two data sets discussed in Section \ref{sec:data}, in units of $\powunits$, measured as a function of wavenumber $k$, and projected for the CO(3-2) line at redshift $z\approx2.5$. Shown separately are the results of our power spectrum analysis of the data from ACA (open circles) and ASPECS (filled circles) data, with the results logarithmically binned over intervals of 0.1 dex. The relative $1\sigma$ uncertainties for each wavenumber bin are also shown for ACA (light gray triangles) and ASPECS (dark gray triangles). For comparison, the $1\sigma$ uncertainties from COPSS are also shown (black triangles), normalized by Equation~\ref{eqn:pow_lineratio} for CO(3-2), using $r_{3,1}=0.42$ \citep{Daddi2015}. Also shown for comparison are Model A and Model B of \citep{Pullen2013}, which have been similarly normalized to CO(3-2). Model C has been generating using the results of Section~\ref{ssec:copss}, setting $\langle T_{b}\rangle^{2}b^{2}=100\ \mu\textrm{K}^2$ and $P_{\rm shot}=2000\ \powunits$ for CO(1-0), and is included for illustrative purposes only.
    \label{fig:mmime_powspec}}
\end{figure*}

\begin{figure}
    \centering
    \includegraphics[width=0.470\textwidth]{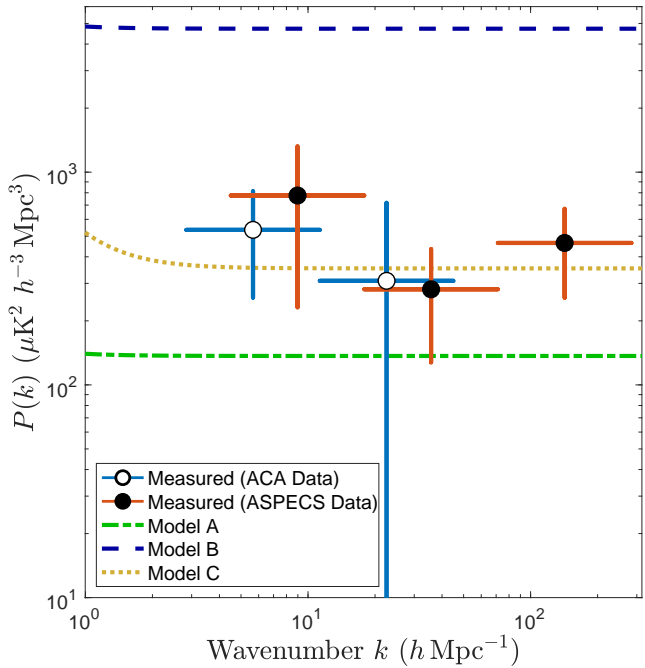}
    \caption{Power measured as a function of wavenumber $k$, projected for the CO(3-2) line at redshift $z\approx2.5$. The points have been binned on a logarithmic interval of 0.6 dex (compared to the 0.1 dex binning used in Figure~\ref{fig:mmime_powspec}).
    \label{fig:mmime_powspec_rebin}}
\end{figure}

Power spectrum results are shown in Figure~\ref{fig:mmime_powspec} and Figure~\ref{fig:mmime_powspec_rebin}, and are reported alongside those from the jackknife analysis in Table~\ref{table:powspec_results}. 

We measure $\tilde{I}^{2}_{s}(\nu)=730\pm240\ \lineunits$ for the ASPECS data. Projecting to the reference frame appropriate for CO(3-2), we find $P_{\textrm{CO(3-2)}}=370\pm120\ \powunits$, with greatest sensitivity between $k=20{-}100\ h\,\textrm{Mpc}^{-1}$. Over this range, we expect the measurement to be completely dominated by shot power, with negligible contributions from large-scale structure.

For the ACA data, we measure $\tilde{I}^{2}_{s}(\nu)=890\pm440\ \lineunits$.  Projecting our results for to that for CO(3-2), we find $P_{\textrm{CO(3-2)}}=460\pm230\ \powunits$, with greatest sensitivity between $k=4{-}20\ h\,\textrm{Mpc}^{-1}$. Similar to the results from ASPECS, we expect this measurement to have negligible contributions from large-scale structure. 

Combining both datasets and weighting by the inverse instrument noise variance, we find $\tilde{I}^{2}_{s}(\nu)=770\pm210\ \lineunits$. With this result, we are able to reject the zero-power hypothesis ($\tilde{I}^{2}_{s}(\nu)=0$) to 99.99\% confidence, signifying that the measured power is unlikely to arise from noise-like fluctuations in the data alone.

\subsection{The Impact of Source Linewidths}\label{sec:disc_linewidths}
For large values of $\eta$, we expect to resolve the lines of individual galaxies, which will suppress the power measured. Our choice to exclude modes where $\eta > 500$ was made in part to allow for a consistent comparison between the results presented here and those from \citetalias{Keating2016}. However, as discussed in \ref{app:linewidths}, this choice will at least partially resolve galaxies with linewidths of $\gtrsim300\ \textrm{km}\ \textrm{sec}^{-1}$. As such objects have been previously observed at high redshift (e.g., \citealt{Tacconi2013,Walter2014,GonzalezLopez2019}), we verify here our assumption that such effects are not affecting our measurements.

A detailed accounting of this effect - and an estimate for a potential correction factor to our measurement - would require modeling that is well beyond the scope of the work presented here.  However, a simpler validation test is to evaluate our results considering a smaller range for the allowed values of $\eta$. Therefore, we recalculate our results using only those modes where $\eta < 250$, which ought to substantially reduce these source-resolving effects for galaxies with linewidths of $\gtrsim300\ \textrm{km}\ \textrm{sec}^{-1}$. If galaxies of linewidth $\sim600\ \textrm{km}\ \textrm{sec}^{-1}$ -- the maximum linewidth found for a galaxy in the ASPECS data \cite{GonzalezLopez2019} -- dominate our measurement, then we expect the power measured with this cutoff to increase by approximately $50\%$.

Including only modes were $\eta < 250$, we measure spectral shot power of $1040\pm620\ \lineunits$ for the ACA data, and $770\pm340\ \lineunits$ for the ASPECS data, reflecting a modest increase of $\sim10\%$ in power. This result is consistent with our prior assumption that our measurement is not dominated by galaxies with broad linewidths of $\gtrsim600\ \textrm{km}\ \textrm{sec}^{-1}$. Though we have not attempted any correction for these effects, we note that if one were to assume that our measurement is solely dominated by objects with linewidths of $300\ \textrm{km}\ \textrm{sec}^{-1}$, we would need to scale up the results presented here by approximately a third.
\subsection{The Impact of Sample Variance}\label{ssec:cv}
\begin{figure}[ht]
    \centering
    \includegraphics[width=0.47\textwidth]{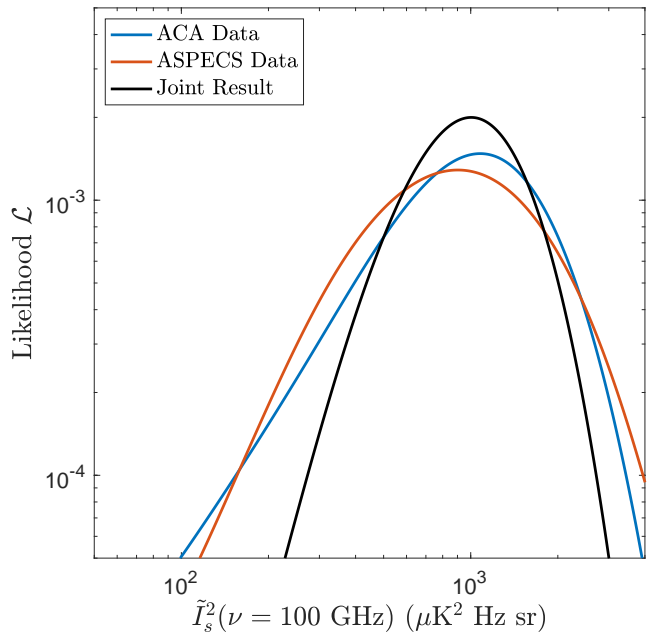}
    \caption{Likelihood estimates of the ``true'' spectral shot power at 100 GHz, accounting for both the effects of instrument noise and sample variance. The likelihood function from the ASPECS data alone (orange) produce a better constraint on the lower end of the range of estimated spectral shot power relative to that from the ACA data (blue), but a poorer constraint on the upper end. This skew of the ACA likelihood function arises both from the decreased significance of sample variance, and the increased instrumental noise. The joint constraint (black) using both data sets is more symmetric, roughly consistent with a log-normal distribution in $\tilde{I}^{2}_{s}(\nu)$.
    \label{fig:linepow_cv}}
\end{figure}

\begin{figure*}[t]
    \centering
    \includegraphics[width=0.975\textwidth]{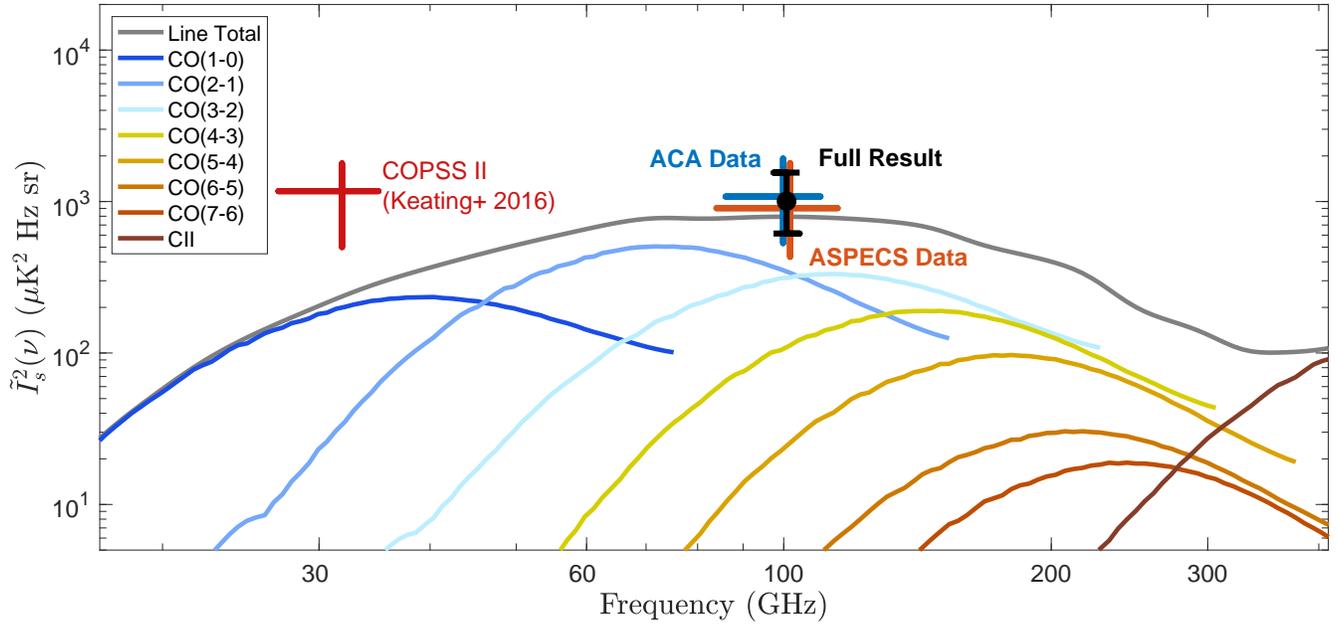}
    \caption{Model estimates for the spectral shot power $\tilde{I}^{2}{s}(\nu)$ -- for both individual line species and the cumulative total -- as a function of frequency, alongside our measurement-based estimates for the spectral shot power (from Section~\ref{ssec:cv}). Constraints from the ACA (blue) and ASPECS (orange) data are shown separately, as well as the constraints from the results of our joint analysis for both datasets (black). We find good agreement between the predicted and measured values. Also shown are the constraints from COPSS (red), after correcting for the estimated contributions from clustering power in the measurement (as discussed in Section~\ref{ssec:copss}). For illustrative effect, we have also included an estimate for the [CII], by using a combination of the \citetalias{Li2016} model, as well as the model $\mathbf{m_{1}}$ from \citet{Silva2015}.
    \label{fig:multi_linepow}}
\end{figure*}

The values and uncertainties in Table~\ref{table:powspec_results} account only for instrumental noise. However, we do expect there to be some intrinsic variance for the relatively small volumes surveyed due to cosmic variance -- imprinted by the inhomogeneity of large-scale cosmic structure -- and Poisson variance --  arising from the discrete number of galaxies within a given volume. We seek here to quantify the effects of both, referring to the combined effects of sample and cosmic variance as ``sample variance'', in order to produce a more accurate estimate (and uncertainties) of the true value of $\tilde{I}^{2}_{s}(\nu)$ at 100 GHz.

For this analysis, we use simulation results of Keenan et al. (2020; submitted), which utilizes the TNG300 simulations from the public data release of Illustris \citep{Nelson2019} to produce simulated power spectra from ensembles of simulated light cones with realistic large-scale structure. From mock observations that approximately match the ACA and ASPECS data sets, we calculate the distribution of measured shot powers within 1000 light cones, normalized by the ``true'' shot power, which we estimate by taking the mean of 1000 large (500 arcmin$^2$) light cones. Fixing the relative contributions of the individual lines in our measurement (estimates of which are discussed further in Section~\ref{ssec:model}), we construct a probability density functions (PDF) for measuring a given amount of power in a field of a given size, relative to the ``true'' spectral shot power. We then run a maximum likelihood estimation (MLE) analysis to calculate the true spectral shot power, given the measurements and uncertainties from the analysis of the ACA and ASPECS data sets. The results of this analysis are reported in Table~\ref{table:comb_results}.

\begin{deluxetable}{c|c|c|c}[t]
\tablecaption{Spectral Shot Power Measurements and Uncertainties\label{table:comb_results}}
\tablehead{
\colhead{Result}   & \colhead{Inst. Noise}   & \multicolumn{2}{c}{Full Analysis}  \\
\colhead{ } & \colhead{Only} & \colhead{\textsc{Median}} & \colhead{\textsc{MLE}}}
\startdata
ACA    & $\phn890_{-440}^{+440}$ & $\phn940_{-500}^{+740}$ & $1080_{-550}^{+850}$\\
ASPECS & $\phn730_{-240}^{+240}$ & $\phn850_{-450}^{+850}$ & $\phn904_{-480}^{+890}$\\
Joint  & $\phn770_{-210}^{+210}$ & $\phn960_{-370}^{+530}$ & $1010_{-390}^{+550}$\\
\enddata
\tablecomments{All values are in units of $\lineunits$.}
\end{deluxetable}

The full likelihood functions from our MLE analysis are shown in Figure~\ref{fig:linepow_cv}, and results are reported in Table~\ref{table:comb_results}. Combining the results of both data sets together, and accounting for both sample and cosmic variance, we estimate the spectral shot power at 100 GHz to be $\tilde{I}^{2}_{s}(\nu)=1010_{-390}^{+550}\lineunits$. This ``most likely'' power is higher than the measured power because, as found in Keenan et al. (2020; submitted), the median power measured in small fields is below the true power due to the skewed distribution of power per field. This result is shown on Figure~\ref{fig:multi_linepow}, alongside the model predictions described in Section~\ref{ssec:model}. We find that, compared to the set of results that consider the instrument noise only as a source of error, the estimates for spectral shot power have not significantly changed, although the error bars have increased by roughly a factor of two. This further underscores the importance of accounting for sample variance in the measurements presented here. As the sample variance is closer to log-normal distribution than a normal one, and as it is the dominant source of uncertainty in our measurement, we have evaluated the likelihoods in Figure~\ref{fig:linepow_cv} on logarithmic intervals of $\tilde{I}^{2}_{s}(\nu)$. We additionally calculate results for the spectral power in Table~\ref{table:comb_results} using the median of the derived likelihood functions, and with this method we find $\tilde{I}^{2}_{s}(\nu)=960_{-370}^{+530}\ \lineunits$, only marginally different than our MLE-derived value.

Finally, we note that in the above analysis, we have assumed that all detected power originates from blindly-detected sources, such that no consideration of selection biases is required. However, as the ACA observations targeted the AzTEC-3 source, we note that there may be an excess of CO(5-4) emission measured within the ACA survey volume. As our model predictions for the contribution of this line are low, and the CO(5-4) line AzTEC-3 source contributes only $\sim1\%$ of the total measured power, we conclude that the effects this potential bias are negligible. The CO(6-5) line of AzTEC-3 is also partially within our frequency coverage and contributes $<1\%$ of the total measured power.

\section{Discussion}\label{sec:disc}
\subsection{Modeling the multiple transitions of CO}\label{ssec:model}
As there is some ambiguity to which line we can ascribe the power measured at 3 mm, we create a simple model, so as to help estimate the amount of power arising from each line species. We adopt the fiducial model of \cite{Li2016}, hereafter referred to as \citetalias{Li2016}, with one significant modification to the conversion between infrared and CO luminosity, which is nominally expressed as
\begin{equation}\label{eqn:ir_to_co}
\log_{10}(L_{IR})= a \log_{10}(L^{\prime}_{\rm CO}) + b,
\end{equation}
where $L_{\rm IR}$ is the bolometric infrared luminosity in units of $L_{\sun}$, and $L^{\prime}_{\rm CO}$ is the line of the object in units of K km sec$^{-1}$ pc$^{2}$. Rather than using the fiducial values from \citet{Carilli2013}, which is only fit against CO(1-0) luminosities (with $a=1.37$, $b=-1.74$), we instead adopt the ``full sample'' correlations from \citet{Kamenetzky2016}, which provide fits for the full set of CO rotational lines that may contribute to our measurement. All other parameters match the fiducial model of \citetalias{Li2016}. Star formation rates (SFR) for individual halos are calculated using the model found in \cite{Behroozi2013}, assuming a logarithmic scatter of 0.3 dex for the star formation rate of individual halos, which are converted into infrared luminosities assuming
\begin{equation}
L_{\rm IR}=10^{10} \left( \frac{\textrm{SFR}}{  M_{\odot}\cdot\textrm{yr}^{-1}} \right ) L_{\odot}.
\end{equation}
CO luminosities for individual halos are assumed to have logarithmic scatter of 0.3 dex around the scaling relation in Equation~\ref{eqn:ir_to_co}. With the exception of halo mass to SFR correlation, each of the aforementioned scaling relationships are assumed to be redshift invariant, which means that the redshift evolution of our model is solely determined by the star formation rates, as calculated by \citeauthor{Behroozi2013}. Although the connection between star formation and molecular gas is a fairly well-studied phenomenon (e.g., \citealt{Schmidt1959,Kennicutt1998}), the redshift evolution of molecular gas is still poorly constrained (Keenan et al. 2020; submitted). As the uncertainties in our measurement are significant, we do not expect this to significantly bias the interpretation of our result.

The model estimates of the individual line contributions are shown in Figure~\ref{fig:multi_linepow}, over a broad window between 15 GHz and 400 GHz, which approximately spans the range of existing and future measurements targeting CO emission \citep{Crites2014,Bower2015,Keating2016,Cleary2016,Lagache2018}. The model predicts a peak of approximately $800\ \lineunits$ at 100 GHz that includes contributions from multiple low-$J$ lines from the epoch of peak star formation \citep{Madau2014}. Our model agrees with the measurement presented here to within 20\% ($\approx0.4\sigma$), as shown in Figure~\ref{fig:multi_linepow}).

We have also tabulated the spectral shot power of individual line species over the frequency interval of the ACA and ASPECS data (Table~\ref{table:model_results}). These values can then be translated into to estimates of the shot power of individual lines by a minor modification to Equation~\ref{eqn:linepow_def}, where
\begin{equation}
P_{\rm shot,line}(z) \approx f_{\rm total} X(z)^{2} Y(z,\nu_{\rm rest,line})\tilde{I}^{2}_{s}(\nu).
\end{equation}

Over the frequency interval of our measurement, our model predicts that the CO(2-1) and CO(3-2) rotational lines will contribute approximately equal amounts of power. The power contribution of the CO(4-3) lower by roughly two-thirds compared to the lower-$J$ lines, while CO(5-4) is lower by about an order of magnitude. Higher-$J$ lines are expected to only contribute a marginal amount of power relative to these three line species, with the sum of all contributions where $J_{\rm upper}\geq6$ equal to less than 1\% of the predicted power. We will therefore neglect the contributions of these lines, and assume that only CO(2-1), CO(3-2), CO(4-3), and CO(5-4) emission are present in our measurement. We also find that these estimates are in rough agreement with the the luminosity function fits derived from direct detection measurements made with ASPECS \citep{Uzgil2019,Decarli2019}, with respect to CO(2-1) and CO(3-2) contributing the bulk of the spectral shot power, followed by CO(4-3).

In our estimates of $f_{\rm total}$, we have excluded the $\textrm{CO(1-0)}$ line from consideration. While the estimated shot power of this line at 100 GHz relatively low, where $P_{\textrm{shot,CO(1-0)}}\sim1\ \powunits$, the spectral shot power it contributes is much more significant, with $\tilde{I}^{2}_{\rm s}(\nu)=230\ \lineunits$. The reason for the apparent discrepancy lies in the normalization term $X^{2}Y$ in Equation~\ref{eqn:linepow_def}, which shrinks to zero as the redshift approaches zero. This feature is a natural consequence of the inverse square law, where relatively faint objects at small luminosity distances can have similar integrated line fluxes as a brighter object at greater distance. For large-area surveys, CO(1-0) would likely be a significant part of the measured spectral shot power. However, for both the ASPECS and ACA datasets, we would expect that the objects dominating the spectral shot power would have line luminosities of $L^{\prime}_{\textrm{CO(1-0)}}\sim10^{9}-10^{10}\ \textrm{K}\ \textrm{km}\ \textrm{s}^{-1}\ \textrm{pc}^2$, which would be readily detected in imaging of both sets of data. As no such sources are found, we conclude that our power spectrum measurement does not have a significant contribution from CO(1-0), and have thus excluded it from our analysis for the sake of clarity. 

\begin{table}[t]
\caption{Model Estimates of Spectral Shot Power at 100 GHz\label{table:model_results}}
\par\smallskip
\begin{center}
\begin{tabular}{c | c | c c | c}
\hline \hline
Line    & $\langle z \rangle$ & $P_{\rm shot}(z)$ & $\tilde{I}^{2}_{s}(\nu)$ & $f_{\rm total}$ \\
Species &                     & $[\powunits]$ & $[\lineunits]$ & \\
\hline 
CO(2-1) & 1.3                 & $100$     & $340$     & $0.43$\\
CO(3-2) & 2.5                 & $160$     & $310$     & $0.40$\\
CO(4-3) & 3.6                 & $\phn80$  & $120$     & $0.14$\\
CO(5-4) & 4.8                 & $\phn20$  & $\phn30$  & $0.03$\\
\hline
Total   & ---                 & ---       & $790$     & $1.00$\\
\hline
\multicolumn{5}{l}{\textsc{Note}--- $f_{\rm total}$ is the fraction of the spectral shot power.}
\end{tabular}
\end{center}
\end{table}

\subsection{Power Spectrum Estimates of Individual Lines}\label{ssec:line_estimates}
\begin{deluxetable*}{ccccc|cc|cc}
\tablecaption{Estimates of Shot Power for Various CO Line Species
\label{table:line_estimates}}
\tablehead{
  \colhead{Line Species} & \colhead{$\langle z \rangle$} & $k_{\rm min}^a$ & $k_{\rm max}^b$ & \colhead{$V_{\rm surv}^c$} & \colhead{$\tilde{I}^{2}_{s}(\nu)$} & \colhead{$P_{\rm shot}(z)$} & \colhead{$P_{\textrm{shot,CO(1-0)}}(z)$} \\
  \colhead{ } & \colhead{ } & \colhead{$[h\,\textrm{Mpc}^{-1}]$} & \colhead{$[h\,\textrm{Mpc}^{-1}]$} & \colhead{$[10^{3}\ \textrm{Mpc}^{3}]$} & \colhead{$[\lineunits]$} & \colhead{$[\powunits]$} & \colhead{$[\powunits]$}
}
\startdata
CO(2-1) & 1.3 & 6 & 230 & 22 & $430_{-170}^{+230}$         & $120_{-50\phn}^{+70\phn}$        & $\phn205_{-90\phn}^{+140}$ \\
CO(3-2) & 2.5 & 4 & 160 & 39 & $400_{-150}^{+220}$         & $210_{-80\phn}^{+110}$           & $1140_{-500}^{+870}$       \\
CO(4-3) & 3.6 & 3 & 130 & 49 & $150_{-60\phn}^{+80\phn}$   & $100_{-40\phn}^{+50\phn}$        & $\phn950_{-440}^{+800}$    \\
CO(5-4) & 4.8 & 3 & 120 & 54 & $\phn30_{-10\phn}^{+20\phn}$& $\phn20_{-10\phn\phn}^{+10\phn}$ & $\phn440_{-200}^{+350}$    \\
\enddata
\tablenotetext{a}{Minimum wavenumber $k$ for which our measurements have meaningful sensitivity, defined as being within a factor of 0.5 dex ($\sim3$) of the peak sensitivity of our power spectrum at a given redshift.} 
\tablenotetext{b}{Maximum wavenumber $k$ for which our measurements have meaningful sensitivity.}
\tablenotetext{c}{Total volume surveyed, summed across both ACA and ASPECS observations.}
\end{deluxetable*}

To better compare with previously published results, we estimate the individual contributions of each line to $\tilde{I}^{2}_{s}(\nu)$, based on the results of our the modeling efforts discussed in Section \ref{ssec:model}. More specifically, we have used our derived values of $f_{\rm total}$ (reported in Table~\ref{table:model_results}) to estimate the power of each individual lines. Results from this analysis are reported in Table~\ref{table:line_estimates}, which show the estimated shot power for each line to be of order $\sim100\ \powunits$. With these values, we now seek to estimate the shot power of the CO(1-0) line at a range of redshifts, using an appropriate set of line ratios, $r_{J,1}$, where we assume that
\begin{equation}\label{eqn:pow_lineratio}
P_{\textrm{shot,CO(}J\textrm{-}J-1\textrm{)}} = r_{J,1}^{2} P_{\textrm{shot,CO(1-0)}},
\end{equation}
where $J$ refers to the upper-level rotational state of CO. We adopt here the set is presented in \citet{Daddi2015}, which used measurements of optically-selected (BzK) galaxies at $z=1.5$ to estimate the relative line ratios of ``normal'' star-forming galaxies at high-redshift. The derived line ratios for $J=2,3,4,5$ are $r_{J,1}=0.76{\pm}0.09,\,0.42{\pm}0.07,\,0.31{\pm}0.06,\,0.23{\pm}0.04$, respectively\footnote{The value for $r_{4,1}$ is based on an interpolation originally performed in \citet{Decarli2016}}. The resulting estimates are shown in Table~\ref{table:line_estimates}. We have used the \citeauthor{Daddi2015} line ratios in place of those from \cite{Kamenetzky2016}, as the former is derived from a sample of high-redshift objects, whereas the latter is derived from low-redshift galaxies. We find similar results when using the ``non-ULIRG'' $L_{\rm IR}$ to $L_{\rm CO}^{\prime}$ correlation from \citeauthor{Kamenetzky2016} in our model, with agreement of order 10-20\% between the two sets of line ratios.

\subsection{Comparison to COPSS II}\label{ssec:copss}
In \citetalias{Keating2016}, it was assumed that all power measured was attributable to shot power only, as several models (e.g., \citealt{Visbal2010,Lidz2011,Pullen2013}) had suggested that the contributions of large-scale structure in the power spectrum would be negligible over the spatial scales that COPSS was most sensitive to ($k=0.5{-}2\ h\, \textrm{Mpc}^{-1}$). However, more recent models (e.g., \citetalias{Li2016}) and our own simulations (Keenan et al. 2020; submitted) indicate that the clustering contribution to the COPSS power should not be ignored.

\begin{figure}[t]
    \centering
    \includegraphics[width=0.47\textwidth]{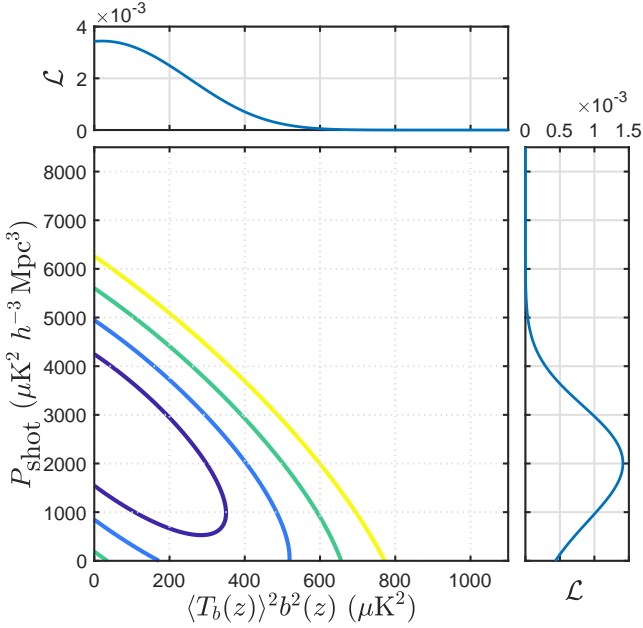}
    \caption{The $\chi^{2}$ statistic for our power spectrum fit of the COPSS CO(1-0) power spectrum measurement, along with the marginalized likelihoods for the individual power spectrum components for $P_{\rm shot}$ and $\langle T_{b}(z)\rangle^{2}b^{2}(z)$. \textit{Center}: The results of the joint fit for the shot power of CO(1-0) and $\langle T_{b}(z)\rangle^{2}b^{2}(z)$ are shown, with contours outlining $\Delta\chi^{2}=1,2.3,4,6.2$ relative to the minimized $\chi^{2}$ value. There is significant covariance between the fits for the two parameters, which are a result of a relatively small number of bins at $k<1\ h\,\textrm{Mpc}^{-1}$ constraining the cluster power component of the power spectrum. \textit{Right:} the marginalized likelihood for $P_{\rm shot}$. As a result of our analysis, we provide a refined estimate for CO(1-0) of $P_{\textrm{shot}}=2000_{-1200}^{+1100}\ \powunits$ at $z=2.6$. We note that if we set $\langle T_{b}(z) \rangle^{2}b^{2}(z)=0$, we recover the original \citetalias{Keating2016} estimate of $3000 \powunits$ \textit{Top:} The marginalized likelihood for $\langle bT \rangle^{2}$. Due to the limited signal-to-noise within the COPSS measurement, we are only able to put an upper limit for CO(1-0) of $\langle T_{b}(z) \rangle^{2} b^{2}(z) <420\ \mu\textrm{K}^{2}$ at $z=2.6$ (95\% confidence). With an appropriate choice of halo-mass function \citep{Tinker2008,Tinker2010}, and estimate for $L(M)$ for CO \citep{Li2016}, we can translate this into an upper limit for the brightness temperature of the CO(1-0) line at $z=2.6$ of $T_{b}<8.2\ \mu\textrm{K}^{2}$ (95\% confidence).
    \label{fig:copss_chi2}}
\end{figure}

Rather than attempting to use approximate astrophysical models, we can use the linear matter power spectrum itself to estimate the contributions from large-scale structure. The power spectrum, $P(k,z)$, as measured in an intensity mapping analysis can be written as
\begin{equation}\label{eqn:genpowspec}
P(k,z) = \langle T_{b}(z) \rangle^{2}b^{2}(z)P_{\textrm{lin}}(k,z) + P_{\textrm{shot}}(z),
\end{equation}
where $P_{\textrm{lin}}$ is the linear matter power spectrum, $b(z)$ is the luminosity-weighted halo bias, and $\langle T_{b}(z) \rangle$ is the mean brightness temperature of the line being examined at a given redshift. In Equation~\ref{eqn:genpowspec}, the $\langle T_{b}(z)\rangle^{2}b^{2}(z) P_{\textrm{lin}}(k,z)$ component is the power arising from the induced clustering of CO emitters from large-scale structure. We further define the halo bias as
\begin{equation}\label{eqn_halobias}
b(z) = \frac{\int^{\infty}_{M_{\textrm{min}}}dM\,L(M)\frac{dn}{dM}b(M,z)}
{\int^{\infty}_{M_{\textrm{min}}}dM\,L(M)\frac{dn}{dM}},
\end{equation}
where $b(M,z)$ is the mass-dependent halo bias, $L(M)$ is the luminosity for a halo of mass $M$. Halos with masses below the low-mass limit, $M_{\textrm{min}}$, are assumed to be deficient in line emission. With a choice of linear power spectrum appropriate for the redshift range of COPSS \citep{Lewis2000,Murray2013}, we solve for the individual components of Equation~\ref{eqn:genpowspec} using the COPSS power spectrum, fitting for $P_{\rm shot}$ and $\langle T_{b}(z) \rangle^{2}b^{2}(z)$ for CO(1-0). We perform a likelihood analysis for both parameters, requiring only that both components be positive, and evaluating the $\chi^{2}$ value for a given set of parameters against the power spectrum presented in \citetalias{Keating2016}. The results of this analysis are shown in Figure~\ref{fig:copss_chi2}.

Marginalizing over the $\langle T_{b}(z) \rangle^{2}b^{2}(z)$ term -- which we have only weak constraints for -- we estimate the shot power of CO(1-0) at $z=2.6$ to be $P_{\rm shot}(z)=2000_{-1200}^{+1100}\ \powunits$. We have shown this result alongside those from our analysis of the ACA and ASPECS data set in Figure~\ref{fig:multi_linepow}. We note that the estimated shot power of the COPSS measurement exceeds the model prediction by nearly a factor of 5. However, the error bars on the COPSS measurement are large, such that the discrepancy between model and measurement is approximately $1.4\sigma$, which is of limited statistical significance.

While not shown in Figure~\ref{fig:multi_linepow}, we note the discrepancy between COPSS and our model is virtually eliminated when using the $L_{\rm IR}-L^{\prime}_{\rm CO}$ scaling relationship derived from the ``non-ULIRG'' sample from \citet{Kamenetzky2016}, although it does raise the expected spectral shot power at 100 GHz by approximately a factor of 3, which creates tension with our current measurement.

\subsection{Comparison to other Blind Surveys for CO Emission}\label{ssec:aspecs_coldz}
We now compare our result to other blind line detection surveys. Using the luminosity functions presented in \citet{Decarli2019}, \citeauthor{Uzgil2019} estimate the spectral shot power at 100 GHz to be $\tilde{I}^{2}_{s}(\nu)=610^{+1040}_{-330}\times10^{2}\ \lineunits$. This values is in good agreement with those derived here, noting that while the measurement presented in this paper is greater by roughly 50\%, both are within the limits of their respective uncertainties. We note that our result is in moderate tension with the power spectrum results of \citet{Uzgil2019}, who found $\tilde{I}^{2}_{s}(\nu)=-930\pm610\ \lineunits$ (excluding the results for $k\geq60\ h\,\textrm{Mpc}^{-1}$, for reasons discussed in Appendix \ref{app:image_powspec}). The difference is larger than what one might expect, as both analyses utilize the ASPECS data set. However, we note that two results were generated using separate reduction pathways, including different flagging schemes and power spectrum estimators. Neglecting corrections for sample variance effects, we find our results from the ASPECS data set are reasonably close to the estimated minimum found in \citeauthor{Uzgil2019} by summing the power contributions of all sources detected at high confidence (shown in Table 2 of \citealt{GonzalezLopez2019}), where $\tilde{I}_{s}^{2}(\nu)>500\ \lineunits$, and similarly find our results (prior to correction for sample variance effects) are below their $3\sigma$ power spectrum upper limit, where $\tilde{I}^{2}_{s}(\nu)<890\ \lineunits$.

The Plateau de Bure High-z Blue Sequence Survey 2 (PHIBSS2; \citealt{Lenkic2020}) provides another set of serendipitous galaxy detections. A calculation of the estimated spectral shot power from the PHIBSS2 source catalog is difficult due to the highly uneven depths of individual pointings and the wide spread of frequencies over which the data were taken. However, as a comparison by proxy, we note that the measured number density of bright CO(3-2) objects appears to be fairly well matched to the derived CO(1-0) luminosity function in \citetalias{Keating2016}, assuming $r_{3,1}=0.42$. That both surveys see such objects may be a byproduct of the relatively large sky areas covered by both measurements, and thus less sensitive to the effects of sample variance.

Though less direct than the comparison with ASPECS, we also compare our results here to those from the CO Luminosity Density at High-z (COLDz) survey \citep{Pavesi2018}. In analyzing the fitted luminosity functions from COLDz, \citet{Uzgil2019} estimate the CO(1-0) shot power at $z=2.4$ to be $P_{\rm shot,CO}=270 \powunits$, which in marginal agreement with what we estimate in Section~\ref{ssec:line_estimates}. This may be the result of sample variance between the fields surveyed (Keenan et al. 2020; submitted).

\subsection{Constraints on Cosmic Molecular Gas Abundance}
\begin{figure*}[t]
    \centering
    \includegraphics[width=0.75\textwidth]{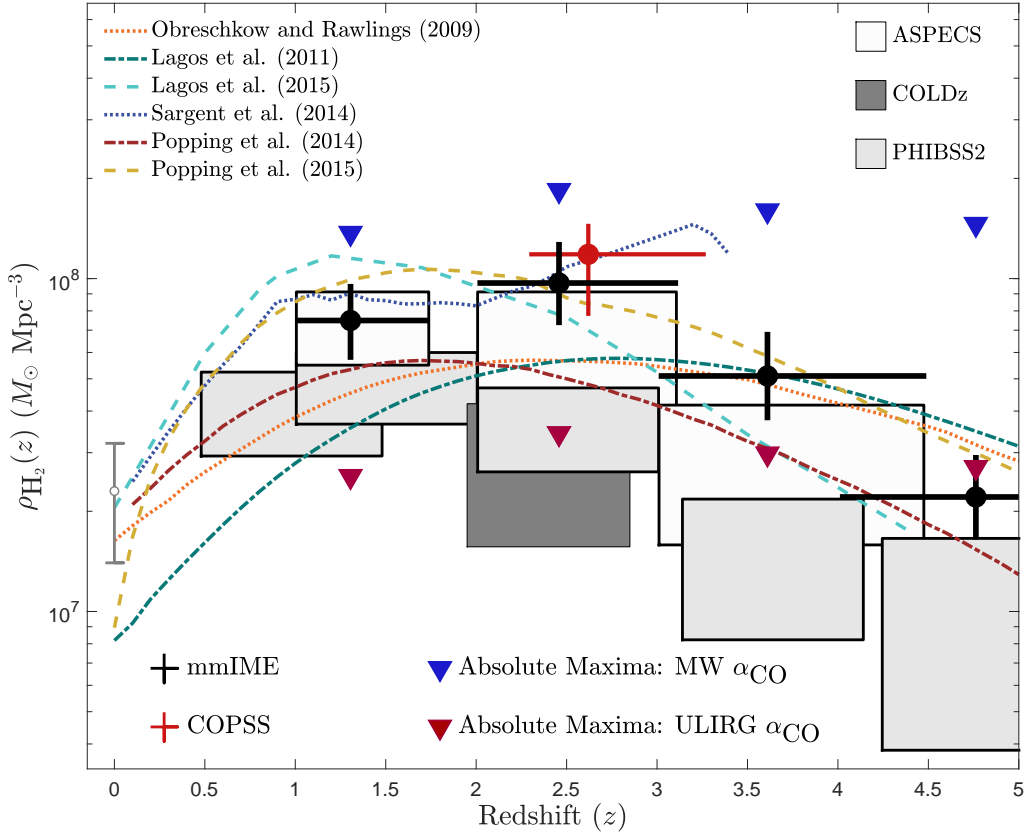}
    \caption{Constraints on the cosmic molecular gas density as a function of redshift. Results from the analysis presented here (black) are in good agreement with those from COPSS (red) and ASPECS (white boxes), although they reside somewhat higher than estimates from PHIBSS2 (light gray boxes) and COLDz (dark gray boxes). A constraint at $z\approx0$ (open circle) from \cite{Keres2003} is also shown. For each redshift bin, we show the maximum power for that redshift assuming that all measured power comes from that redshift bin, using the Milky Way conversion between luminosity and gas mass (see text; blue triangles), the same assumption made to place the black points, and the same limit assuming a ULIRG-like conversion (red triangles). For comparison, an ensemble of theoretical models for $\rho_{H_{2}}(z)$ are shown: \cite{Obreschkow2009c} is shown as dotted orange, \cite{Lagos2011} is shown as dot-dashed green, \cite{Lagos2015} is shown as dashed teal, \cite{Sargent2014} is shown as dotted blue, \cite{Popping2014} is shown as dot-dashed brown, and \cite{Popping2015} is shown as dashed yellow. The constraints presented here are in broad agreement with this ensemble of models, although constraints at $z=2.5$ reside at the top end of the range of predictions.
    \label{fig:rho_h2}}
\end{figure*}
\begin{deluxetable}{ccc|cc}[t]
\tablecaption{Estimates of the Cosmic Molecular Gas Density
\label{table:rhoh2_estimates}}
\tablehead{
  \colhead{$\langle z \rangle$} & \colhead{Redshift} & \colhead{$\rho_{\textrm{H}_2}(z)$} & \multicolumn{2}{c}{Absolute Maximum$^a$} \\
  \colhead{ } & \colhead{Range} & \colhead{$[10^{7}\ M_{\sun}\ \textrm{Mpc}^{-3}]$} & \multicolumn{2}{c}{$[10^{7}\ M_{\sun}\ \textrm{Mpc}^{-3}]$} \\
  \colhead{ } & \colhead{ } & \colhead{ } & \colhead{\scriptsize{\textsc{All MW$^b$}}} & \colhead{\scriptsize{\textsc{All ULIRG$^c$}}}
}
\startdata
1.3 & $1.0{-}1.7$ & $\phn7.5_{-1.8}^{+2.2}$ & $13.7$ & $2.5$ \\
2.5 & $2.0{-}3.1$ & $\phn9.7_{-2.5}^{+3.2}$ & $18.4$ & $3.4$ \\
3.6 & $3.0{-}4.5$ & $\phn5.1_{-1.3}^{+1.8}$ & $16.0$ & $3.0$ \\
4.8 & $4.0{-}5.8$ & $\phn2.2_{-0.6}^{+0.7}$ & $14.8$ & $2.7$ \\
\hline
2.6 & $2.3{-}3.3$ & $11.8_{-4.1}^{+2.8}$    & ---    & ---   
\enddata
\tablenotetext{a}{Assumes all emission arises from a single redshift window.}
\tablenotetext{b}{Uses $\alpha_{\rm CO}=4.3\ M_{\sun}\ (\textrm{K}\ \textrm{km}\ \textrm{s}^{-1}\ \textrm{pc}^{2})^{-1}$}
\tablenotetext{c}{Uses $\alpha_{\rm CO}=0.8\ M_{\sun}\ (\textrm{K}\ \textrm{km}\ \textrm{s}^{-1}\ \textrm{pc}^{2})^{-1}$}
\end{deluxetable}
Having derived estimates for the CO(1-0) shot power at a range of redshifts, we now seek to translate these into constraints on the cosmic molecular gas density, $\rho_{\textrm{H}_{2}}(z)$. To do so, we note that several works have found that the fractional mass of molecular gas within a given halo, $f_{\textrm{H}_2}$, appears to peak at $M_{\rm halo}\approx10^{12}\ M_{\sun}/h$ \citep{Lagos2011,Popping2015,Lagos2015}, approximately the virial mass of the Milky Way \citep{Watkins2019} -- which appears to be redshift-invariant. For halos below this mass limit, $f_{\textrm{H}_2}$ is estimated to scale in near-linear fashion with halo mass. Based on these theoretical findings, we estimate that the CO(1-0) luminosity of a given halo can be approximated by
\begin{eqnarray}\label{eqn:halomass_to_lum}
L_{\textrm{CO}}(M) &=& A_{\textrm{CO}} \frac{M^{2}}{M_{0}};\ M\leq M_{0} \\
L_{\textrm{CO}}(M) &=& A_{\textrm{CO}} M_{0};\ M\geq M_{0}.
\end{eqnarray}
In Equation~\ref{eqn:halomass_to_lum}, $M$ is the halo mass, $L_{\textrm{CO}}(M)$ is the expected CO(1-0) luminosity for a halo of a given mass, $A_{\textrm{CO}}$ is the mass to CO(1-0) luminosity ratio for halos of mass $M_{0}$, where we have set $M_{0}=10^{12}\ M_{\sun}/h$. In effect, the luminosities of halos scale as $M^{2}$ up to $M_{0}$, above which halos are assumed to have approximately equal luminosity. This scaling relationship is similar to that found in \citetalias{Li2016}, as well as that effectively assumed for our adapted model in Section~\ref{ssec:model}.

Using the fiducial values from \citetalias{Li2016} for the scatter between halo mass and CO luminosity, and utilizing the halo mass functions of \citealt{Tinker2008,Tinker2010}, we are able to fit for the value of $A_{\textrm{CO}}$ using the CO(1-0) estimates using the \citeauthor{Daddi2015} estimates for the CO line ratios. Assuming a value $\alpha_{\rm CO}=3.6\ M_{\sun}\ (\textrm{K}\ \textrm{km}\ \textrm{s}^{-1}\ \textrm{pc}^{2})^{-1}$, we are able to translate the CO luminosities from Equation~\ref{eqn:halomass_to_lum} into molecular gas masses, and integrating over all halo masses, are able to provide an estimate for $\rho_{\textrm{H}_{2}}(z)$. Results from this analysis are reported in Table~\ref{table:rhoh2_estimates}, and shown in Figure~\ref{fig:rho_h2}.

As there is some ambiguity as to how much emission is truly arising from each redshift window, we calculate an additional set of estimates, where we assume that \emph{all} of the power measured originates from a single spectral line at a given redshift, using the \citeauthor{Daddi2015} line ratios to generate an estimate for CO(1-0). With these absolute maxima, we consider two separate scenarios to further assist in our interpretation of the data. For the first, which we refer to as the  ``All MW'' scenario, we adopt the Milky Way value for $\alpha_{\rm CO}$, where $\alpha_{\rm CO,MW}=4.3\ M_{\sun}\ (\textrm{K}\ \textrm{km}\ \textrm{s}^{-1}\ \textrm{pc}^{2})^{-1}$ \citep{Frerking1982,Dame2001,Bolatto2013}. In the second, which we refer to as the ``All ULIRG'' scenario, we adopt the value of $\alpha_{\rm CO}$ found from observations of local ULIRGs, where $\alpha_{\rm CO,ULIRG}=0.8\ M_{\sun}\ (\textrm{K}\ \textrm{km}\ \textrm{s}^{-1}\ \textrm{pc}^{2})^{-1}$ \citep{Downes1998}. These results are also reported in Table~\ref{table:rhoh2_estimates}, and shown in Figure~\ref{fig:rho_h2}.

Comparing our results to theoretical estimates, we find that the mmIME data points are comparable to the theoretical models, except at $z\approx2.5$, where mmIME is higher than all but one model. Our maxima from the all-MW estimates lie well above the ensemble of models. This would appear to place a cap of $\rho{\textrm{H}_{2}}(z)\lesssim2\times10^{8}\ M_{\sun}\ \textrm{Mpc}^{-3}$ across all of cosmic time.

Comparing our results against those previous experiments, we find that our estimates lie above those from COLDz and PHIBSS2, being more consistent with ASPECS. We find very good agreement between mmIME, ASPECS, and COPSS at $z\approx2.5$, with all three values clustered around $10^{8}\ M_{\sun}\ \textrm{Mpc}^{-3}$. The discrepancy with the PHIBSS2 data could result from incompleteness in their variable-depth survey. \citet{Lenkic2020} do not account for sources below their detection threshold, which may include $L^{\prime *}$ galaxies at $z>2$. We also find that COLDz estimates are also well below those presented here. 

We particularly note the large spread in results at $z\approx2{-}3$, where the present measurements (and their associated $1\sigma$ error intervals), stretch over an order of magnitude. That the intensity mapping-derived constraints lie above those from direct detection suggest that such measurements may be capturing emission from galaxies below direct detection threshold. This possibility is explored in Keenan et al., (2020; submitted), and is a serious concern of \citet{Popping2019}.

\section{Conclusion}\label{sec:conclusion}
We have presented the first results from mmIME, an intensity mapping experiment targeting cool gas tracers in the millimeter-wave regime. We have utilized data designed for conducting blind surveys at 3 mm, over an approximate total survey volume of $10^{5}\ \textrm{Mpc}^{3}$, finding the following results:
\begin{itemize}
  \item We report a measurement of $730_{-240}^{+240}\ \lineunits$ (99.9\% confidence) in our analysis of the ASPECS 3-mm data set \citep{GonzalezLopez2019}, and a measurement of $890_{-440}^{+440}\ \lineunits$ (97.8\% confidence) in our analysis of an ACA 3-mm data set collected on a region within the COSMOS field. Accounting for sample variance effects, we estimate the spectral shot power at 100 GHz to be $\tilde{I}^{2}_{s}(\nu)=1010_{-390}^{+550}\ \lineunits$, rejecting the zero-power hypothesis ($\tilde{I}^{2}_{s}(\nu)=0$) to 99.99\% confidence. We find that these results are in good agreement with an expanded version of a model first presented in \citet{Li2016}, using the $L_{\rm IR}{-}L^{\prime}_{\rm CO}$ scaling relationships measured in \citet{Kamenetzky2016}.
  \item Using the aforementioned model, we have produced estimates of the shot power at 100 GHz for CO(2-1), CO(3-2), CO(4-3), and $\textrm{CO(5-4)}$ of $P_{\rm shot}$ =  $120^{+80}_{-40}$, $200^{+120}_{-70}$, $90^{+70}_{-40}$, and $20^{+10}_{-10}\ \powunits$, respectively.
  \item Using a set of line ratios appropriate for normal, Milky Way-like galaxies at high redshift \citep{Daddi2015}, we have produced estimates of the CO(1-0) shot power at $z{=}1.3$, $2.5$, and $3.6$. We find good agreement between these values and a refined estimate of the shot power from COPSS of $2000^{+1100}_{-1200}\ \powunits$.
  \item We set constraints on the cosmic molecular gas density between $z\approx1{-}5$, finding $\rho_{\textrm{H}_{2}}(z)$ = $7.5_{-1.8}^{+2.2}\times10^{7}$, $9.7_{-2.5}^{+3.2}\times10^{7}$, and $5.1_{-1.3}^{+1.8}\times10^{7}\ M_{\sun}\ \textrm{Mpc}^{-3}$ for $z{=}1.3$, $2.5$, and $3.6$, respectively. For $\rho_{\textrm{H}_{2}}(z{\approx}3)$, we find good agreement with the results of COPSS and ASPECS.
\end{itemize}

While tantalizing, the results presented here will likely require further observational follow-up, particularly as the uncertainties on our measurement are significant. Fortunately, further constraints from intensity mapping experiments targeting CO are on the horizon. The SMA is presently engaged in intensity mapping-focused observations as part of mmIME, targeting aggregate CO emission in the 1.3 mm atmospheric window. The ASIAA Intensity Mapping of CO (AIM-CO; \citealt{Bower2015}) project will similarly target lines in the 3 mm window, with prospects for cross-correlation with measurements from COPSS. The CarbON CII line in post-rEionization and ReionizaTiOn epoch project (CONCERTO; \citealt{Lagache2018}) will, among other things, make power spectrum measurements at 2 mm and 1.3 mm, likely probing a similar a set of lines similar to what we have examined here. Similarly, the Tomographic Ionized-carbon Mapping Experiment (TIME; \citealt{Crites2014}), while targeting [CII] emission at $z\approx6{-}10$, will also be capable of measuring CO in the 1.3-mm atmospheric window. Finally, the CO Mapping Array Pathfinder (COMAP; \citealt{Cleary2016}) will measure the CO(1-0) line at $z\approx2.5$, with access to much larger spatial scales than were originally measured in COPSS. 

There is significant potential for intensity mapping experiments of CO and [CII] for probing the astrophysics of the gas fueling the star formation within early galaxies. Numerous theoretical studies have evaluated the potential of such experiments for placing constraints on cosmology (e.g., \citealt{Fonseca2017,Bernal2019}), covering a variety of topics from the expansion history of the early Universe \citep{Karkare2018} to the physics of inflation \citep{Moradinezhad2019b}. Such experiments are likely to benefit from the decades of significant technical and observational studies of the CMB, and may be highly compatible with future experiments targeting the CMB (e.g, \citealt{Switzer2017,Moradinezhad2019a,Delabrouille2019}). 

\acknowledgments

The authors would like to thank the anonymous referee for their timely and thoughtful feedback, which helped improve the quality of this manuscript. We thank T. Li for his contributions and collaboration in this earliest stages of this work. We thank S. Paine for his assistance in identifying atmospheric features, and F. Fornasini for her feedback on several figures prepared for this manuscript. We thank N. Mashian, T.-C. Chang, A. Fialkov, and A. Loeb for the many useful discussions had on theoretical aspects related to this work. GKK would additionally like to several members on the intensity mapping community who provided thoughtful feedback and useful conversations over the course of this project, including P. Breysse, P. Bull, D. Chung, A. Crites, A. Dizgah, O. Dor\'e, K. Karkare, A. Pullen, and G. Sun. 

This paper makes use of the following ALMA data: ADS/JAO.ALMA \#2013.1.00146.S, ADS/JAO.ALMA \#2016.1.00324.L, ADS/JAO.ALMA \#2016.1.01149.S, ADS/JAO.ALMA \#2018.1.01594.S. ALMA is a partnership of ESO (representing its member states), NSF (USA) and NINS (Japan), together with NRC (Canada), MOST and ASIAA (Taiwan), and KASI (Republic of Korea), in cooperation with the Republic of Chile. The Joint ALMA Observatory is operated by ESO, AUI/NRAO and NAOJ. The National Radio Astronomy Observatory is a facility of the National Science Foundation operated under cooperative agreement by Associated Universities, Inc. DPM and RPK were supported by the National Science Foundation through CAREER grant AST-1653228. RPK was supported by the National Science Foundation through Graduate Research Fellowship grant DGE-1746060.

\appendix
\section{The Impact of Linewidths and Instrument Resolution}\label{app:linewidths}
The design of intensity mapping experiments, and the interpretation of observational data as three-dimensional power spectra, must consider the impact of internal galaxy motions on the intensity mapping power spectrum. Galaxy rotation, turbulence, and mergers spread the line emission in redshift space over hundreds of $\textrm{km}\ \textrm{sec}^{-1}$. Because radial velocity is mapped to distance through the Hubble flow, this width corresponds to Mpc of radial distance, and therefore an attenuation of power in modes where $k\gtrsim 0.1\ \textrm{Mpc}^{-1}$. As several planned intensity mapping experiments will have at least some sensitivity to to this range in $k$, we consider here the impact of such effects here.

\subsection{The Impact of Realistic Line Emission Profiles}\label{sapp:line_profiles}
\begin{figure}
    \centering
    \includegraphics[width=0.975\textwidth]{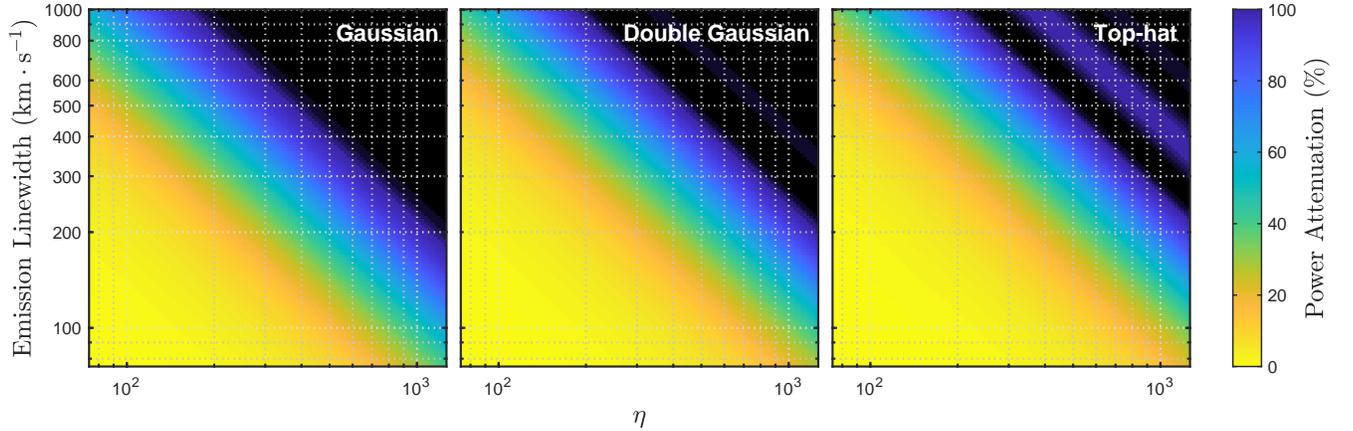}
    \caption{Estimated attenuation of power as a function of emission linewidth and $\eta$. Typically, the power is attenuated by half when $\eta \sim c/2\Delta{v}_{\rm line}$. We note that the attenuation here is shown on a per-halo basis, and thus most appropriate for calculations of the shot power, although similar attenuation will be seen in the clustering component of the power spectrum.  At $\eta\approx130{-}190$ (depending on line profile type), the attenuation in power is $\approx50\%$ for $\sigma_{\rm line}=616\ \textrm{km}\ \textrm{sec}^{-1}$, the maximum linewidth found in \cite{GonzalezLopez2019}. 
    \label{fig:atten_impact}}
\end{figure}

For this analysis, we represent galaxies as three different velocity profiles: top-hat, Gaussian, and double-Gaussian. The flux density for a top-hat line profile is assumed a constant non-zero value over a velocity range defined by the line width, $\Delta{v}_{\rm line}$, and zero outside of this range. For the Gaussian line profile, we set the full width at half maximum (FWHM) of the profile to be $\Delta{v}_{\rm line}$. The double-Gaussian line profile is defined as
\begin{equation}\label{eqn:double_gauss}
S_{\rm line}(\Delta{v}_{\rm line},\Delta{v}) = A\left (e^{((B\Delta{v}-\Delta{v}_{\rm line})/\Delta{v}_{\rm line})^{2}} + e^{((B\Delta{v}+\Delta{v}_{\rm line})/\Delta{v}_{\rm line})^{2}}\right ),
\end{equation}
where $\Delta v$ is the velocity offset from line center, and $A$ and $B$ are normalization constants, where $A$ is set by the line luminosity of the object, and $B=3.6421$. This represents two equal Gaussians of FWHM $\approx\Delta{v}_{\rm line}/2$, spaced roughly $\Delta{v}_{\rm line}/2$ apart. 

Shown in Figure~\ref{fig:atten_impact} is the attenuation for a given linewidth for the three line profiles as a function of wavenumber $\eta$. Generally speaking, modes with significant attenuation reside at $\eta\gtrsim100$, with the Gaussian line profile showing generally more attenuation than the top-hat line profile at a given value of $\eta$ and $\Delta v_{\rm line}$, and the double-Gaussian attenuation generally found between the two. Based on the modeling of \cite{Li2016}, we expect galaxies within halos of $\sim 10^{12}M_{\sun}$ to provide the bulk of the aggregate line emission for CO in particular, and are estimated to have linewidths of $\sim300\ \textrm{km}\ \textrm{sec}^{-1}$ \citep{Obreschkow2009e}, which will be attenuated by 50\% at $\eta\approx450$. Our analysis of the ACA and ASPECS data sets exclude $\eta>500$ modes to reduce the impact of this line attenuation. Integrating across all wavenumbers used in the analysis of the ASPECS and ACA data, we estimate that the band-averaged power is attenuated by $\sim25\%$ for emitters with linewidths of $300\ \textrm{km}\ \textrm{sec}^{-1}$, and $\sim50\%$ for emitters of $600\ \textrm{km}\ \textrm{sec}^{-1}$. These losses drop to $\sim10\%$ and $\sim25\%$ respectively when only including modes where $\eta<250$. Without a model distribution of linewidths, we choose not to make any correction for this effect in our power spectrum.

\subsection{The Impact of Channel Resolution}\label{app:linewidths_chanres}
\begin{figure}[t]
    \centering
    \includegraphics[width=0.975\textwidth]{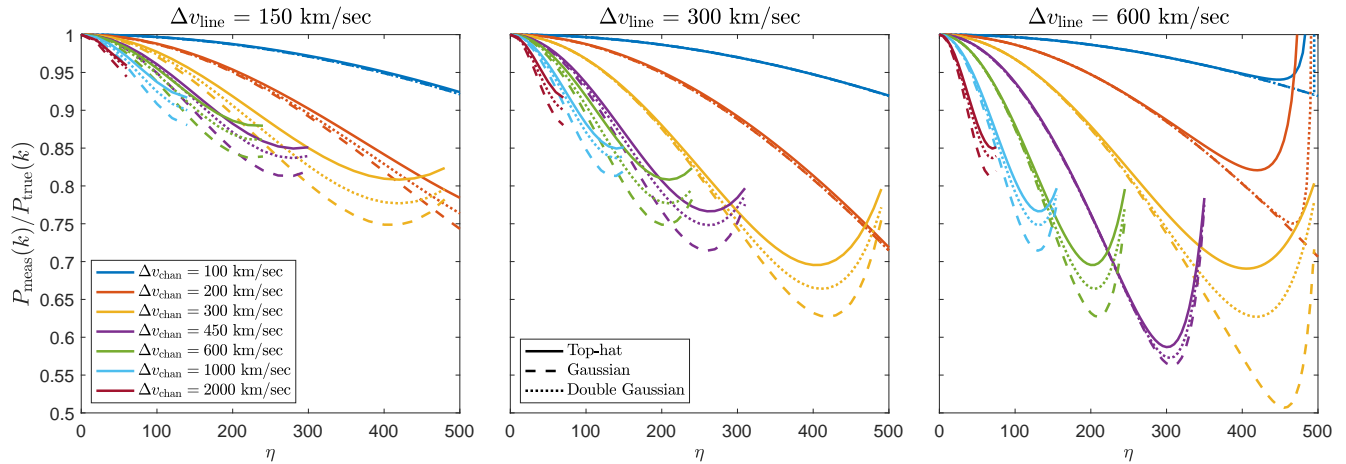}
    \caption{Power recovered in the presence of instrument resolution effects and non-infinitesimal linewidths. In the figures above, power has been averaged over each plane of constant $\eta$, and normalized by estimated power in the absence of channel resolution effects. In the right-most panel, some spectra show a significant increase in power at very high values of $\eta$. This can be attributed to the combination of aliasing and the low power intrinsically present in these modes (as shown in Figure~\ref{fig:atten_impact}).
    \label{fig:chanres_effect}}
\end{figure}

We now turn to the impact of the instrumental spectral resolution on line intensity mapping measurements. We run separate trials using each of the line profiles mentioned above, generating a mock catalog of sources with fixed $\Delta{v}_{\rm line}$, and random sky position and center frequency. We then generate an image cube, simulating the expected emission to be found within each voxel for a given channel resolution, $\Delta{v}_{\rm chan}$, expressed in velocity units.

The results of our analysis are shown in Figure~\ref{fig:chanres_effect}. In general, we find that the recovered power decreases at all values of $\eta$ with increasing $\Delta v_{\rm chan}$. For a band-averaged value across all accessible values of $\eta$, we find that the losses are most significant when $\Delta{v}_{\rm chan} \sim \Delta{v}_{\rm line}$, where the recovered power is $\approx85\%$ of what one would expect given $\Delta{v}_{\rm chan} << \Delta{v}_{\rm line}$. With decreased resolution (i.e., $\Delta{v}_{\rm chan} > \Delta{v}_{\rm line}$), the band-averaged losses in power are somewhat diminished, although can still produce losses of order $\sim5\%$. For the analysis presented in Section~\ref{sec:results} (where $\Delta{v}_{\rm chan} \sim 0.1\Delta{v}_{\rm line}$), we estimate the loss in band-averaged power due to instrument resolution effects to be $\lesssim1\%$.

We note here that we have adopted a simple top-hat function for the spectral PSF, rather than something more closely related to the typical out put on an instrument (e.g., a sinc or sinc$^2$ function for an interferometer). We have done so in part because for the data sets discussed in Section~\ref{sec:data}, the native spectral resolution of the data is quite fine, such that the the width of the PSF is quite small relative to the linewidths in question. In averaging down the data for analysis, our resultant PSF resembles something fairly close to a top-hat function. Accounting for the impact of the spectral PSF is an instrument-specific task, though we note that with few exceptions, most are likely to see similar losses in the high-$\eta$ modes.

\section{Image-domain Power Spectrum Analysis with Interferometers}\label{app:image_powspec}

The power spectrum analysis presented in this paper works in the visibility domain that is the natural output of interferometric measurements. In this space, the Fourier dual of the image domain, the noise is well understood and uncorrelated between measurements. Each visibility also naturally represents a specific Fourier component, which makes the mapping to the power spectrum measurement very simple. 

Of course, interferometric observations are most often converted to image space for scientific interpretation and so it is worth considering the path through the image domain to a power spectrum. In particular, this approach was used in the \cite{Uzgil2019} analysis of the ASPECS data to measure the CO power spectrum. They report $P(k)_{\textrm{CO(2-1)}}=-45\pm77\ \powunits$, while our visibility-domain analysis of the same data finds $P(k)_{\textrm{CO(2-1)}}=190\pm60\ \powunits$ if all power is ascribed to CO(2-1), without correction for sample variance effects. In this appendix we examine the image-domain power spectrum analysis of that paper using the simulation tools developed for our visibility analysis procedure. We expect that a visibility-domain analysis will produce something close to a minimum-variance estimate of the power, and we have extensively evaluated our method for biases using simulations and found it to reliably measure the correct power on average. We are therefore most interested in characterizing any biases or uncertainty differences that arise through the image-domain analysis.

Our simulated data sets are based on the ASPECS LP 3 mm observations. We replace the visibility data with combinations of synthetic noise, with properties derived from the data themselves, and mock galaxy distributions. The mock galaxy catalogs are based on the luminosity function fits found in \cite{Decarli2019}. We assume the sources to be delta function-like in both spectral and spatial domains, so that there is no potential impact from resolving individual sources. The position of each galaxy was randomized for during each trial, with no attempt to mimic clustering or large-scale structure, over a volume approximately 10 times larger than the mosaicked area covered by ASPECS. To simplify interpretation, we adjust the luminosity of all objects in a single mock catalog so that the mean power measured from the full volume is equal to $P(k)=200\ \powunits$ when normalized to the redshift for CO(2-1) ($z\approx1.3$).

Once we have synthetic data containing realistic noise and galaxy distributions, we follow the methodology laid out in \citeauthor{Uzgil2019}. Data are spectrally averaged down by a factor of 40, such that the channel resolution becomes 156.25~MHz. Visibilities for each pointing of the mosaic are split into one of four subsets of data. Data from each subset are then independently imaged using natural weighting and are mosaicked together, with a correction for the attenuation of the primary beam applied to the mosaicked map. A $110"\times110"$ region from the mosaicked dirty map is extracted, which roughly fits within the half-power point of the mosaicked area. Though we have not added continuum sources, we do replicate the filtering process designed to remove them, subtracting of a linear fit across frequency to the emission at given position in the sky. The resultant image cubes are then converted to brightness temperature units (in K, from the more typical Jy/beam found in interferometric images), Fourier transformed, and cross-correlated against one another, with the results of the auto-correlations dropped in order to prevent noise from positively biasing the measurement. Data are converted to power spectrum units of $\powunits$, utilizing the projection appropriate for CO(2-1), after which the data are binned into a 1D power spectrum. Data from the $\eta=0$ plane are not included, as the continuum fitting procedure above will have removed all power from that set of modes. Similarly, data from the column of modes where the $uv$-distance is zero are also removed, as an interferometer should not contain information on the set of ``zero-spacing'' modes.

\begin{figure}[t]
    \centering
    \includegraphics[width=0.975\textwidth]{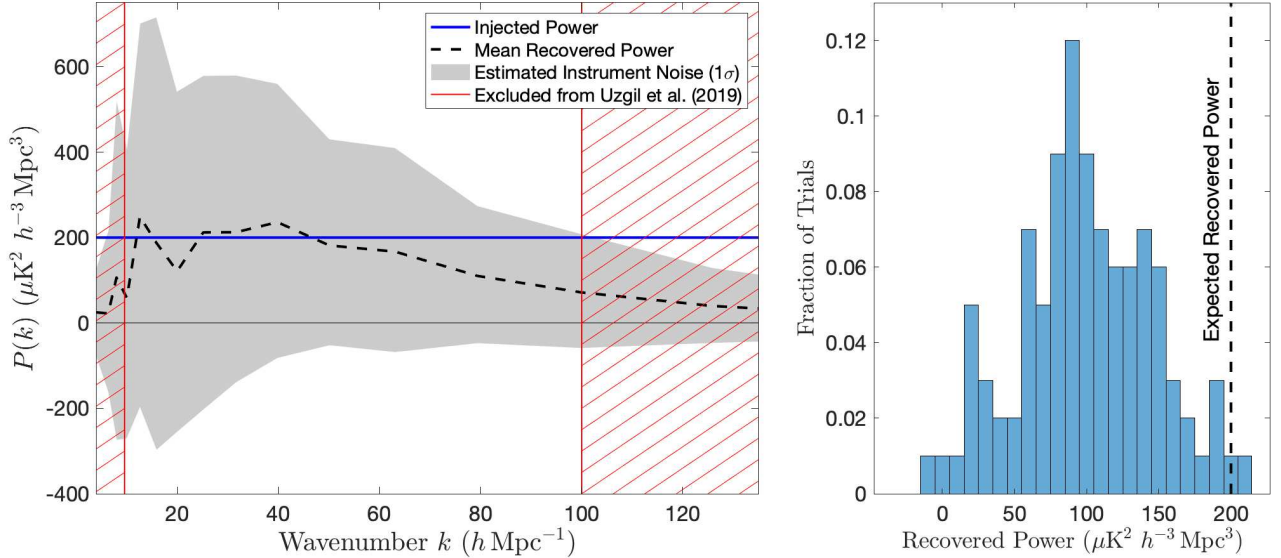}
    \caption{\textit{Left}: The power spectrum recovered using the methodology from \cite{Uzgil2019}. The mean injected power is shown with a blue horizontal line, the mean recovered power with a black dashed line. The gray region represents the 16-84\% range for the 100 independent mock observations. The power spectrum is analyzed with bins of width $4.1\ h\,\textrm{Mpc}^{-1}$, in order to match the bin widths used in \cite{Uzgil2019}. The areas excluded by the analysis of \citeauthor{Uzgil2019} are indicated in red. \textit{Right}: A histogram of band-averaged measurements for each trial, using inverse-variance weighting with incorrectly estimated per-bin variances. For the data shown here, we have also accounted for the $\approx15\%$ loss due to channel resolution effects. The mean recovered power across all trials is approximately a factor of two below the true power in the simulations. 
    \label{fig:singledish_sim}}
\end{figure}

The distribution of power spectra measured in 100 trials of these simulations is shown in Figure~\ref{fig:singledish_sim}. The gray area represents the 16-84\% range, and a similar width is found in noise-only simulations, indicating that the measurement noise dominates over variance in the galaxy catalog. There are significant deviations from the injected power at several wavenumbers, particularly for $k\lesssim10\ h\,\textrm{Mpc}^{-1}$, which roughly corresponds to the shortest baselines present within the dataset. There is also strong attenuation for $k\gtrsim50\ h\,\textrm{Mpc}^{-1}$, exceeding 50\% for $k\gtrsim80\ h\,\textrm{Mpc}^{-1}$. \citeauthor{Uzgil2019} hypothesize that the attenuation is the result of resolving the map on size-scales of the synthesized beam, although at $\approx2^{\prime\prime}$ resolution, one would expect to see such effects at $\approx130\ h\,\textrm{Mpc}^{-1}$, as seen in Figure~\ref{fig:mmime_powspec}. In further testing with different array configurations for the ALMA 12-m array, simulating a range of hour angle coverages, we find that the recovered power for this image-space analysis is highly sensitive to the $uv$-coverage, with less uniform coverage giving rise to more anomalous structure in the recovered power spectrum. This suggests that the source of the apparent structure is the non-uniform nature of the synthesized aperture of the interferometer. While uniform weighting of the data may help to mitigate this (e.g., \citealt{Mertens2020}), the use of natural weighting may actually exacerbate such issues, particularly in the case where the noise performance of different elements of the interferometer varies.

In attempting to derive the most sensitive measurement possible, \cite{Uzgil2019} use inverse-variance weighting to combine the individual bins of the power spectrum to provide a single band-averaged value (as was done in Section~\ref{ssec:powspec_methods} here). Their weighting relies on bin uncertainties derived from the correlation products of jackknifed maps, whose power spectrum noise statistics are expected to be very similar to those shown in Figure~\ref{fig:singledish_sim} (left). Unfortunately, the attenuation of high-$k$ modes affects both signal \emph{and} noise, which leads to an underestimate of the variance in the high-$k$ bins and thus their over-weighting in the overall average. In Figure~\ref{fig:singledish_sim} (right), we show the recovered average power when we replicate this procedure, using the variances derived for each bin from the 100 trials, represented as the gray areas in the left panel. The recovered power is $120\ \powunits$ averaging over $9.55\leq k \leq 100.05\ h\,\textrm{Mpc}^{-1}$ (matching the \citealt{Uzgil2019} range), and $80\ \powunits$ averaging over the full $4.11\leq k \leq 130.9\ h\,\textrm{Mpc}^{-1}$ range of the data. \citeauthor{Uzgil2019} do publish a separate result for the range $9.55\leq k\leq 55.0\ h\,\textrm{Mpc}^{-1}$, which we note is relatively free of the aforementioned effects. We use the value they measure over this interval, $P(k)=-260\pm170 \powunits$, in our discussion of different experiments in Section~\ref{ssec:aspecs_coldz}.

The mock catalogs used to test this method use galaxies with zero linewidth. However, the previous appendix notes the potential for signal loss due to the interplay of the galaxy emission lines and the spectral resolution of the data. The image-domain analysis of \cite{Uzgil2019} average over 40 channels, corresponding to a channel resolution of $\Delta{v}_{\rm chan}\simeq450\ \textrm{km}\ \textrm{sec}^{-1}$, comparable to the linewidths of the source population under study. As discussed in Appendix \ref{app:linewidths}, this results in losses from channel resolution effects to be maximized, which should introduce an additional 15\% loss. After accounting for this effect, we estimate that the recovered power in simulations that contain $200\ \powunits$ will be $100\ \powunits$ averaging over $9.55\leq k \leq 100.05\ h\,\textrm{Mpc}^{-1}$, and $70\ \powunits$ averaging over $4.11\leq k \leq 130.9\ h\,\textrm{Mpc}^{-1}$.

We note that while deconvolution might appear to help resolve the issue raised here by effectively filling in the missing information in the $uv$-domain, it is likely to give rise to problems akin to ``mode-mixing'' \citep{Datta2010,Morales2012}, where power may effectively cast throughout the $(u,v,\eta)$ domain, potentially destroying the spatial scale information one hopes to recover in a power spectrum measurement. As such, methods like CLEAN \citep{Hogbom1974} are unlikely to prove useful with image domain-based intensity mapping analyses.

\bibliography{bibliography.bib}
\end{document}